\documentclass[onecolumn]{aastex631}

\usepackage{graphicx}

\shorttitle{three large amplitude contact binaries}
\shortauthors{Xu et al.}
\graphicspath{{./}{figures/}}
\begin{document}

\title{ Photometric and Spectroscopic Investigations of Three Large Amplitude Contact Binaries }

\author{Xin Xu}
\author[0000-0003-3590-335X]{Kai Li}
\correspondingauthor{Kai Li}
\email{kaili@sdu.edu.cn}
\author{Fei Liu}
\author{Qian-Xue Yan}
\author{Yi-Fan Wang}
\author{Xin-Yu Cui}
\author{Jing-Yi Wang}
\affiliation{Shandong Key Laboratory of Optical Astronomy and Solar-Terrestrial Environment, School of Space Science and Technology, Institute of Space Sciences, Shandong University, Weihai, Shandong 264209, People's Republic of China}

\author{Xing Gao}
\affiliation{Xinjiang Astronomical Observatory, 150 Science 1-Street, Urumqi 830011, People's Republic of China}

\author[0000-0003-3162-3350]{Guo-You Sun}
\affiliation{Xingming Observatory, Urumqi 830002, Xinjiang, People's Republic of China}

\author{Cheng-Yu Wu}
\author{Mu-Zi-Mei Li}
\affiliation{Shandong Key Laboratory of Optical Astronomy and Solar-Terrestrial Environment, School of Space Science and Technology, Institute of Space Sciences, Shandong University, Weihai, Shandong 264209, People's Republic of China}
\begin{abstract}
We performed photometric and spectroscopic studies of three large amplitude contact binaries, NSVS 2418361, ATLAS J057.1170+31.2384 and NSVS 7377875. The amplitudes of three systems' light curves are more than 0.7 magnitude. We analyzed the light curves using Wilson-Devinney code to yield physical parameters. The photometric solutions suggested that NSVS 7377875 belongs to an A-subtype contact binary, while the others are classified as  W-subtype ones. Furthermore, the mass ratio of NSVS 7377875 is higher than 0.72, so it belongs to H-subtype contact binaries. Since their light curves have unequal height at two maxima which is called O'Connell effect,  a dark spot on the primary component for each target was required to get a better fit of light curves. The orbital period investigation shows that the period of NSVS 2418361 is increasing, indicating a mass transfer from the less massive component to the more massive one, while the other targets exhibit no long-term variation. Our spectral subtraction analysis of LAMOST spectra revealed excess emissions in the $H_\alpha$ line, indicating chromospheric activity in all the three targets. The Gaia distance was applied to estimate the absolute parameters of the three targets, and we obtained their evolutionary state. The relationships between the energy transfer parameter of 76 H-subtype contact binaries and their bolometric luminosity ratios, as well as their contact degree, were presented. We discovered that H-subtype systems have less efficient energy transfer rate, which is corresponding to the conclusion proposed by Csizmadia \& Klagyivik.
\end{abstract}

\keywords{Close binary stars (254); Eclipsing binary stars (444); Mass ratio (1012); Fundamental properties of stars (555); Stellar evolution(1599) }

\section{Introduction} 
Close binary systems are classified into three types. A binary system belongs to a detached binary when neither of its components fill its Roche lobe; if only one of the components fills its Roche lobe, the binary belongs to a semi-detached binary; when both components fill their Roche lobes, the binary system is classified as a contact binary. W UMa-type contact binaries are composed  of two dwarf stars sharing a common convection envelope. They usually consist of two late spectral type (F,G,K) stars, and recent studies shows that some M spectral type stars are included \citep{1993,Terrell_2012,Drake_2014}. The light curves of contact binaries exhibit approximately equal depth at primary and secondary minima, indicating that the surface effective temperature of each component is nearly identical. This phenomenon can be explained by the energy transfer occurring in the common convection envelope \citep{Lucy}. Contact binaries are classified into two subtypes, A-subtype and W-subtype. For A-subtype, the more massive component is hotter; for W-subtype, the more massive component is cooler \citep{AW}. \citet{2020MNRAS} revealed that the evolutionary pathways for the two subtypes are different.
\par There is no final conclusion about the formation and evolution of W UMa contact binaries. It is suggested that these systems are evolved from short period detached binaries due to angular momentum loss (AML) provided by the magnetic braking, and will ultimately coalesce into single stars \citep{AML5,AML4,AML1,AML2,AML6,AML3,massratiolimit1}. However, only one contact binary merging event has been observed, that is V1309 Sco \citep{V1309}. Additionally, the theory of thermal relaxation oscillation (TRO) \citep{TRO1,TRO2,TRO3} predicted that the components of contact binaries are not in thermal equilibrium and a contact binary will oscillate between contact-broken phase and contact-phase. This theory requires further observational data and additional samples to verify its predictions. Furthermore, the questions such as short period cut-off \citep{periodcut2,periodcut1,2020AJLi}, mass ratio limit \citep{massratiolimit3,massratiolimit2,2021ApJLi,2023MNRAS.519.5760L,2023A&A...672A.176P} and magnetic activity of W UMa systems are still open issues. Recently, \citet{2023A&A...672A.176P} employed Bayesian inference methods, utilizing data from the Kepler Eclipsing Binary Catalog and Gaia to study the mass ratio limit issue. By analyzing the distribution of light curve amplitudes, they inferred the mass ratio distribution and determined the minimum mass ratio ($q_{min}$) values for  contact binaries with different periods. 
\par The magnetic activity  which related to the convective mass transfer and rapid rotation is common in late-type stars \citep{starspots}. It manifests in various forms, including star spots, flares, and plages \citep{magnetic,magnetic2}. The two components of contact binaries are usually late-type stars, suggesting strong magnetic activity of them. The star spots model can be applied to interpret the asymmetric of light curves which is called O'Connell effect \citep{O'Connell}. What's more, the emission lines of $H_\alpha, H_\beta$, Ca $\mathrm{II}$ H\&K and Ca $\mathrm{II}$ IRT can be served as the indicators of chromospheric activity \citep{Halpha,Ca}.
\par High-mass ratio, shallow contact binaries serve as valuable tools for investigating the evolutionary status of close binaries, particularly their intricate transition from a near-contact to a contact phase \citep{2010PASJQian}. \citet{q=0.72} proposed H-subtype contact binaries which have a mass ratio higher than 0.72. They suggested that the energy transfer behaviors of these contact binaries are unique compared to other contact binaries. However, \citet{Sun2020ApJS}  didn't find the energy transfer behaviors of H-subtype contact binaries they found are special compared to others. Therefore, more samples and studies of high mass ratio contact binaries are needed to verify these statements. \citet{qamplitude} have found that the amplitude grows with mass ratio and the third light is expected to affect the amplitude. Since large amplitude contact binaries are likely to be high mass ratio systems, we selected three contact binaries with amplitude larger than 0.7 magnitude from the variable star catalog of All-Sky Automated Survey for SuperNovae (ASAS-SN; \citealt{2014ASASSN,2023ASASSN}). The previous studies found that for totally eclipsing contact binaries, the mass ratios of photometric ones and the spectroscopic ones are nearly equal \citep{lightcurve, Gaia}. Recently, \cite{2024arXiv240600155R} has found a mass ratio discrepancy between spectroscopic and photometric methods of AW UMa, casting doubt on the widely used Lucy model of contact binaries. Due to the absence of radial velocity curves, we performed photometric studies to determine their physical parameters. All the basic information of our three targets were obtained from the ASAS-SN website\footnote{\url{ https://asas-sn.osu.edu}} and are shown in Table \ref{ASAS-SN}.
\par \citet{2017RMxAA..53..133K} performed photometric study of J084000 and determined its mass ratio is 0.898, so it belongs to H-subtype contact binaries. Recently, \citet{11000} used the machine learning method to study the ASAS-SN V-band light curves of more than 11,200 short-period  EW-type eclipsing binary. They determined the mass ratio of J080043 is 0.73, the mass ratio of J073802 is 0.99. Therefore, both of them can be catalogued to high mass ratio contact binaries initially. For J034828, we conducted the first photometric and spectroscopic study. 

{\begin{deluxetable*}{cccccc}
		\tablecaption{ The basic information of the three targets from ASAS-SN database.\label{ASAS-SN}}
		\tablenum{1}
		\tablehead{\colhead{Target} & \colhead{Hereafter} & \colhead{Other names} & \colhead{Period} & \colhead{V} & \colhead{Amplitude} \\ 
			\colhead{} & \colhead{} & \colhead{} & \colhead{(days)} & \colhead{(mag)} & \colhead{(mag)} } 
		\startdata
		ASASSN-V J034828.11+311418.4 & J034828 & NSVS 2418361  & 0.2527926 & 15.46 & 0.76 \\
		ASASSN-V J073802.91+585021.6 & J073802 & ATLAS J057.1170+31.2384 & 0.3596081 & 14.32 & 0.75 \\
		ASASSN-V J084000.43+363927.9 & J084000 & NSVS 7377875 & 0.2649852 & 13.41 & 0.83 \\ 
		\enddata
		
	\end{deluxetable*} 

\section{Observations} 
\subsection{Photometric Observations}
We conducted photometric observations on the three contact binaries using  Weihai Observatory 1.0-m telescope of Shandong University (WHOT; \citealp{WHOT}),  the 60 cm Ningbo Bureau of Education and Xinjiang Observatory Telescope (NEXT) and the 60 cm telescope at the Xinglong Station of National Astronomical Observatories (XL60). A PIXIS 2048B CCD camera is equipped on WHOT, the scale of each pixel is approximately 0.35 arcsec, resulting a 12 arcmin× 12 arcmin field of view. A back-illuminated FLI 230-42 CCD camera is equipped on NEXT, the camera has a 22 arcmin × 22 arcmin field of view resulted by 2048 × 2048 pixels.  An Andor DU934P camera is equipped on XL60, the field of view is 18 arcmin ×18 arcmin. Based on the C-MUNIPACK\footnote{\url{https://c-munipack.sourceforge.net/}} software, we performed bias, dark, and flat corrections. After performing aperture photometry, we determined the differential magnitudes between the variable stars and the comparison stars, as well as the magnitudes between the comparison stars and the check stars. The observation information and the fundamental information of the comparison and check stars are shown in Table \ref{2MASS}. The photometric observation data of three targets are listed in Table \ref{Photometry}, where $\Delta \text{MAG}$ is the differential magnitudes between the variable stars and the comparison stars. Our three systems have also been observed by various photometric surveys, including Super Wide Angle Search for Planets (SuperWASP; \citealt{SWASP}), Catalina Real-time Transient Survey (CRTS; \citealt{CRTS}), and Zwicky Transient Facility (ZTF; \citealt{ZTF1,ZTF2}). Additionally, the Transiting Exoplanet Survey Satellite (TESS; \citealt{TESS}) observed two of our targets: J034828 was observed by Sectors 42, 43, 44 and 71, while J073802 was observed by Sectors 47, 60 and 74.

\vspace{-1cm}
\begin{deluxetable*}{cccccc}
	
	\tablecaption{The observation information of the three contact binaries.
		\label{2MASS}}
	
	\tablenum{2}
	
	\tablehead{\colhead{Target} & \colhead{Observational date} & \colhead{Exposure times} & \colhead{Comparison star} & \colhead{Check star} & \colhead{Telescope}\\
		\colhead{} & \colhead{} & \colhead{(s)} & \colhead{} & \colhead{} & \colhead{}} 
	
	\startdata
	J034828 & 2023 Dec 06 & R90 I70 & 2MASS 03482708+3115114 & 2MASS 03480923+3115065 & WHOT \\
	J073802 & 2021 Dec 10,11 & g75 r65 i85 & 2MASS 07382661+5847489 & 2MASS 07381593+5854417 & NEXT \\
	J084000 & 2022 Dec 29 & R70 I70 & 2MASS 08400724+3633512 & 2MASS 08395920+3632116 & XL60 \\
	\enddata
	
\end{deluxetable*}

\begin{deluxetable*}{ccccc}
	\tablecaption{The photometric observation of J034828, J073802 and J084000.\label{Photometry}}
	\tablenum{3}
	
	\tablehead{\colhead{Target} & \colhead{HJD-R} & \colhead{$\Delta$ \text{MAG-R}} & \colhead{HJD-I} & \colhead{$\Delta$ \text{MAG-I}} \\ 
		\colhead{} & \colhead{} & \colhead{(mag)} & \colhead{} & \colhead{(mag)} } 
	
	\startdata
	J034828 & 2460284.98431  & 0.262  & 2460284.98318  & 0.114  \\
	& 2460284.98657  & 0.262  & 2460284.98544  & 0.105  \\
	& 2460284.98883  & 0.266  & 2460284.98770  & 0.115  \\
	& 2460284.99109  & 0.278  & 2460284.98996  & 0.108  \\
	& 2460284.99335  & 0.280  & 2460284.99222  & 0.129  \\
	& 2460284.99561  & 0.297  & 2460284.99448  & 0.130  \\
	& 2460284.99787  & 0.300  & 2460284.99674  & 0.131  \\
	& 2460285.00013  & 0.316  & 2460284.99900  & 0.139  \\
	\enddata
	\tablecomments{This table is available in its entirety in machine-readable form in the online version of this article.}
\end{deluxetable*}

\subsection{Spectroscopic Observations}
Large Sky Area Multi-Object Fiber Spectroscopic Telescope (LAMOST, also known as the Guoshoujing Telescop) is a 4-meter reflecting Schmidt telescope. It is equipped with 4,000 fibers evenly distributed across its focal surface to improve the spectral acquisition rate, and it has a field of view of 5\degr \citep{LAMOST}.  Therefore, it has great potential to efficiently survey a large volume of space for stars and galaxies. In its low-resolution mode, the spectral resolution is approximately 1800, covering a wavelength range from 3700 to 9000\AA. J034828, J073802 and J084000 have been observed by LAMOST in low resolution R$\sim$ 1800 and we found six spectra  from Data Release 10\footnote{\url{http://www.lamost.org/dr10/v1.0/}}.  The observational date, signal-to-noise ratio of g filter (SNRg), the spectral type, radial velocity (RV) and the atmospheric parameters including the effective temperature ($T_{eff}$), surface gravity (log g) and metallicity abundance ([Fe/H]) are listed in Table \ref{tab:spectrum}. We also queried the atmospheric parameters from {\it Gaia} DR3 \citep{Gaia2016,Gaia2023} and listed them in Table \ref{tab:spectrum}. The parameters especially the temperatures derived from Gaia are compared with those obtained from the LAMOST. For J073802 and J084000, the results are quite similar. However, for J034828, a temperature difference of 400 K was observed.

\begin{deluxetable*}{cccccccccc}
	\tablecaption{The atmosphere parameters  of the three binaries provided by LAMOST and Gaia.
		\label{tab:spectrum}}
	\tablenum{4}
	\tablehead{\colhead{Target} & \colhead{Source of data} & \colhead{Observational date} & \colhead{Exposures} & \colhead{SNRg} & \colhead{$T_{eff}$} & \colhead{log g} & \colhead{[Fe/H]} & \colhead{Radial velocity} & \colhead{EW}\\ 
		\colhead{} & \colhead{} & \colhead{} & \colhead{(s)} & \colhead{} & \colhead{(K)} & \colhead{(cgs)} & \colhead{} & \colhead{$(\text{kms}^{-1})$}& \colhead{(\AA) } }
	\startdata
	J034828 & LAMOST & 2016/10/28 & 1800 & 62 & 4914$\pm$24 & 4.325$\pm$0.034 & -0.049$\pm$0.021 & -1.14$\pm$3.21 & 0.741±0.028 \\
	& Gaia & -- & -- & -- & 5356$\pm$33 & 4.443$\pm$0.009 & 0.097$\pm$0.037 &-- & -- \\ 
	J073802 & LAMOST & 2016/12/3 & 1800 & 102 & 5690$\pm$27 & 4.304$\pm$0.037 & 0.198$\pm$0.021& -32.91$\pm$3.82& 0.746±0.054 \\
	& LAMOST & 2016/12/3 & 1800 & 124 & 5662$\pm$37 & 4.342$\pm$0.050 & 0.173$\pm$0.028 & -47.76$\pm$5.65 & 0.681±0.064 \\
	& Gaia & -- & -- & -- & 5528$\pm$9 & 4.288$\pm$0.017 & 0.453$\pm$0.032 & 50.01$\pm7.76$ & -- \\ 
	J084000 & LAMSOT & 2013/1/29 & 1800 & 38 & 4872$\pm$24& 4.311$\pm$0.032 & -0.267$\pm$0.022 & 43.17$\pm$3.12 & 0.899±0.086 \\
	& LAMOST & 2014/12/8 & 1800 & 24 & 4836$\pm$39 & 4.459$\pm$0.061 & -0.331$\pm$0.041 & 34.71$\pm$3.66 & 0.873±0.041 \\
	& LAMOST & 2014/12/26 & 1800 & 79 & 4877$\pm$32 & 4.465$\pm$0.045 & -0.273$\pm$0.027 & 42.65$\pm$4.32 & 0.811±0.041 \\
	& Gaia & -- & -- & -- & 4927$\pm$25 & 4.449$\pm$0.010 & -0.346$\pm$0.055 &-- & -- \\  
	\enddata
	
\end{deluxetable*}

\section{Light Curve Analysis}
Using the 2013 version of the Wilson-Devinney (W-D) code \citep{wd1,wd2,wd3}, we analyzed the light curves of the three targets. The following parameters were adjustable: the orbital inclination (i); the phase shift; the effective temperature of star 2 ($T_2$); the monochromatic luminosity of star 1 ($L_1$); the dimensionless potential ($\Omega = \Omega_1 = \Omega_2$).  Since a temperature difference of 400 K was observed for J034828, we utilized the deredden color indices to calculate the temperatures. The color indices \textit{B - V} and \textit{g - r} were sourced from the AAVSO Photometric All Sky Survey (APASS) Data Release 10 \citep{APASS}, while the \textit{J - K} indices were obtained from the Two Micron All Sky Survey (2MASS; \citealt{2MASS}). The extinction coefficients were obtained from the IRSA\footnote{\url{https://irsa.ipac.caltech.edu/applications/DUST/}} database, then the dereddened color indices were calculated. Ultimately, we inferred the temperature based on the table of \citet{BCv}. The results are listed in Table \ref{5}. For J034828, the temperatures inferred from the color indices are more consistent with the LAMOST temparature rather than Gaia temperature. For the other targets, the temperatures derived from color indices, LAMOST and Gaia are nearly equal. Therefore, it's reliable to employ the average LAMOST temperature as the initial effective temperature of the primary component in the W-D program.
\begin{deluxetable}{ccccccc}
	\tablecaption{Temperatures of the three targets.
	\label{5}}

	\tablenum{5}

	\tablehead{\colhead{Target} & \colhead{$T_{\text{LAMOST}}$(K)} & \colhead{$T_{Gaia}$(K)} & \colhead{$T_{B-V}$(K)} & \colhead{$T_{J-H}$(K)} & \colhead{$T_{g-r}$(K)} & \colhead{$T_{\text{color indices}}$(K)}  }
	\startdata
	J03 & 4914 & 5356 & 4967 & 4858 & 4759 & 4861 \\
	J07 & 5676 & 5528 & 5910 & 5412 & 5198 & 5543 \\
	J08 & 4861 & 4926 & 5216 & 4860 & 4371 & 4816 \\
	\enddata
	\tablecomments{$T_{\text{LAMOST}}$ is the average value of the LAMOST temperature for each target. $T_{\text{color indices}}$ is the average temperature determined from color indices for each target.}

\end{deluxetable}
 \par Given that the $T_{eff}$ of all the systems are lower than 7200K, indicating they have convective envelopes, we set the gravity-darkening coefficients as $g_1 = g_2 = 0.32$ \citep{Lucy1967}, bolometric albedo coefficients as $A_1 = A_2 = 0.5$ \citep{A1A2}. The bolometric and bandpass limb-darkening coefficients were interpolated from Van Hamme's table \citep{Table} with a square root law ($ld = -3$).
Owing to the absence of radial velocity curves for our targets, we employed the q-search method to determine the mass ratio ($q$ = $m_2/m_1$). In this process, We set the step size as 0.1. The relationships between the mean residuals and the mass ratio $q$ for all the systems are displayed in Figure \ref{qsearch}. As shown in the figure, the mass ratio with minimum mean residuals can be obtained, then we set it as the initial value and a free parameter to obtain a convergent solution. We also set the third light ($l_3$) as an adjustable parameter to run the W-D program. The results of $l_3$ were either zero or negative, so we concluded that the third light is not present.  Notably, all light curves of our observations exhibits unequal magnitudes at phases 0.25 and 0.75, a phenomenon known as O'Connell effect. To fit the asymmetric light curves, we added a cool spot on the primary or secondary component of each target.  For J034828, J073802, and J084000, the mean residuals of adding spots on the primary and secondary components are 0.00209 and 0.00215, 0.00085 and 0.00089, and 0.00143 and 0.00153, respectively.  Since adding the spot on the primary component results in smaller mean residuals, we selected this option. Subsequently, utilizing the temperature ratio ($T_{2i}/T_{1i}$) and radius ratio ($k$ = $r_2/r_1$) obtained from the W-D code, we employed the following equations \citep{T1T2} to compute the updated temperature of each component:
\begin{eqnarray}
	T_1 &= \left(\frac{(1 + k^2)T_{eff}^4}{1 + k^2 \left(\frac{T_{2i}}{T_{1i}}\right)^4}\right)^{\frac{1}{4}} \nonumber \\
	T_2 &= T_1 \left(\frac{T_{2i}}{T_{1i}}\right).
\end{eqnarray}
 %图一
\begin{figure}
	\epsscale{0.36}
	\plotone{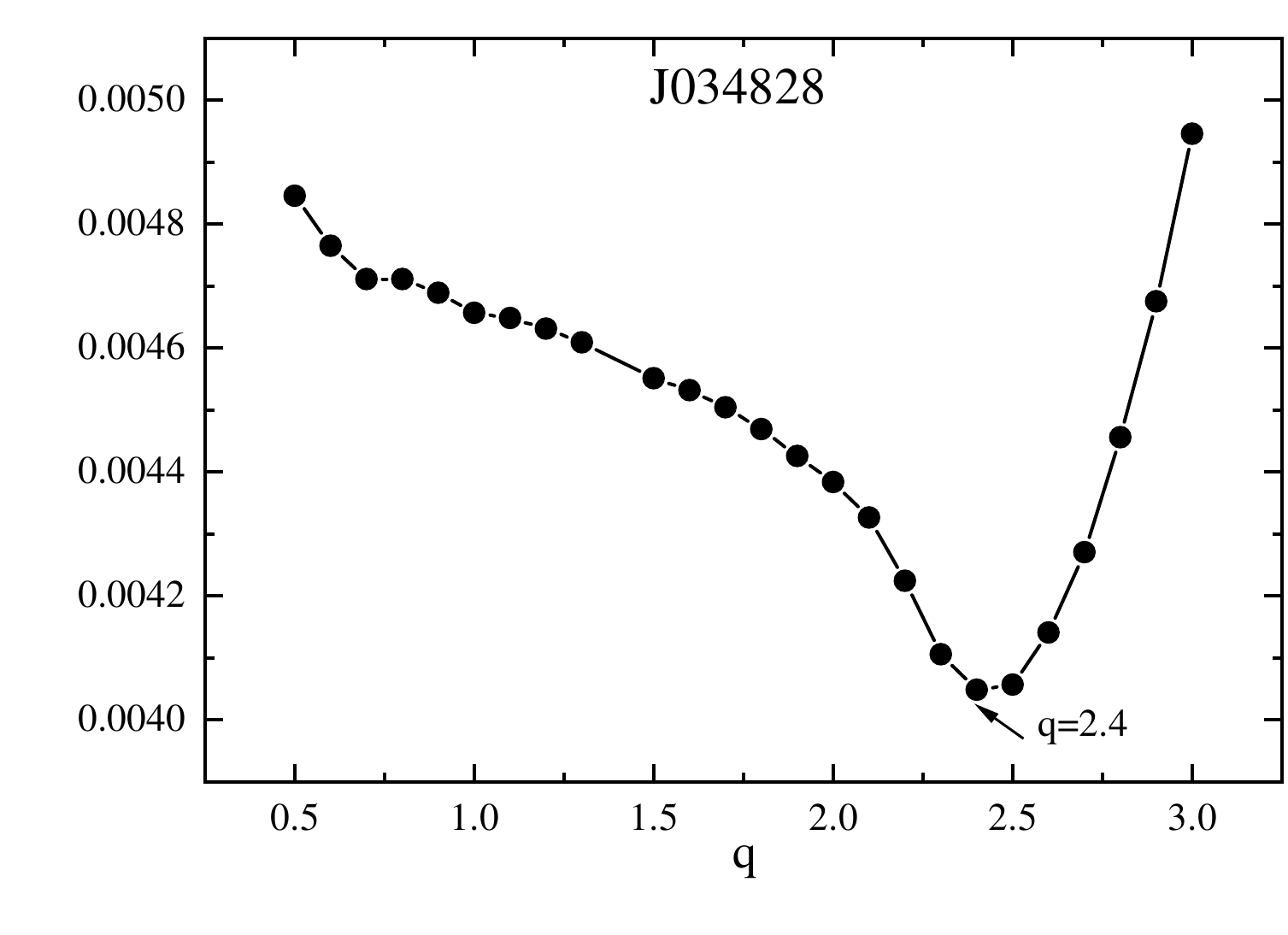}
	\plotone{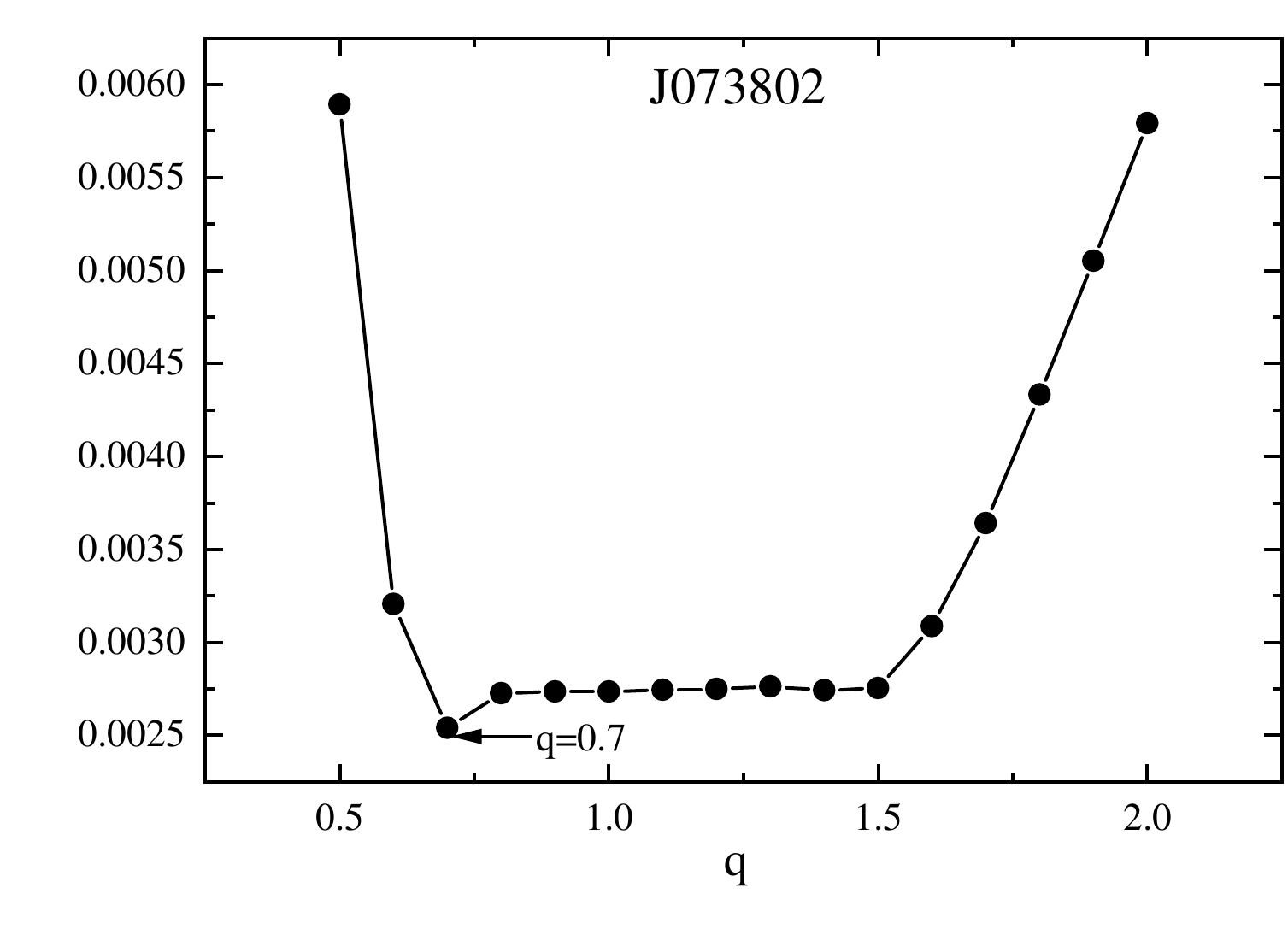}
	\plotone{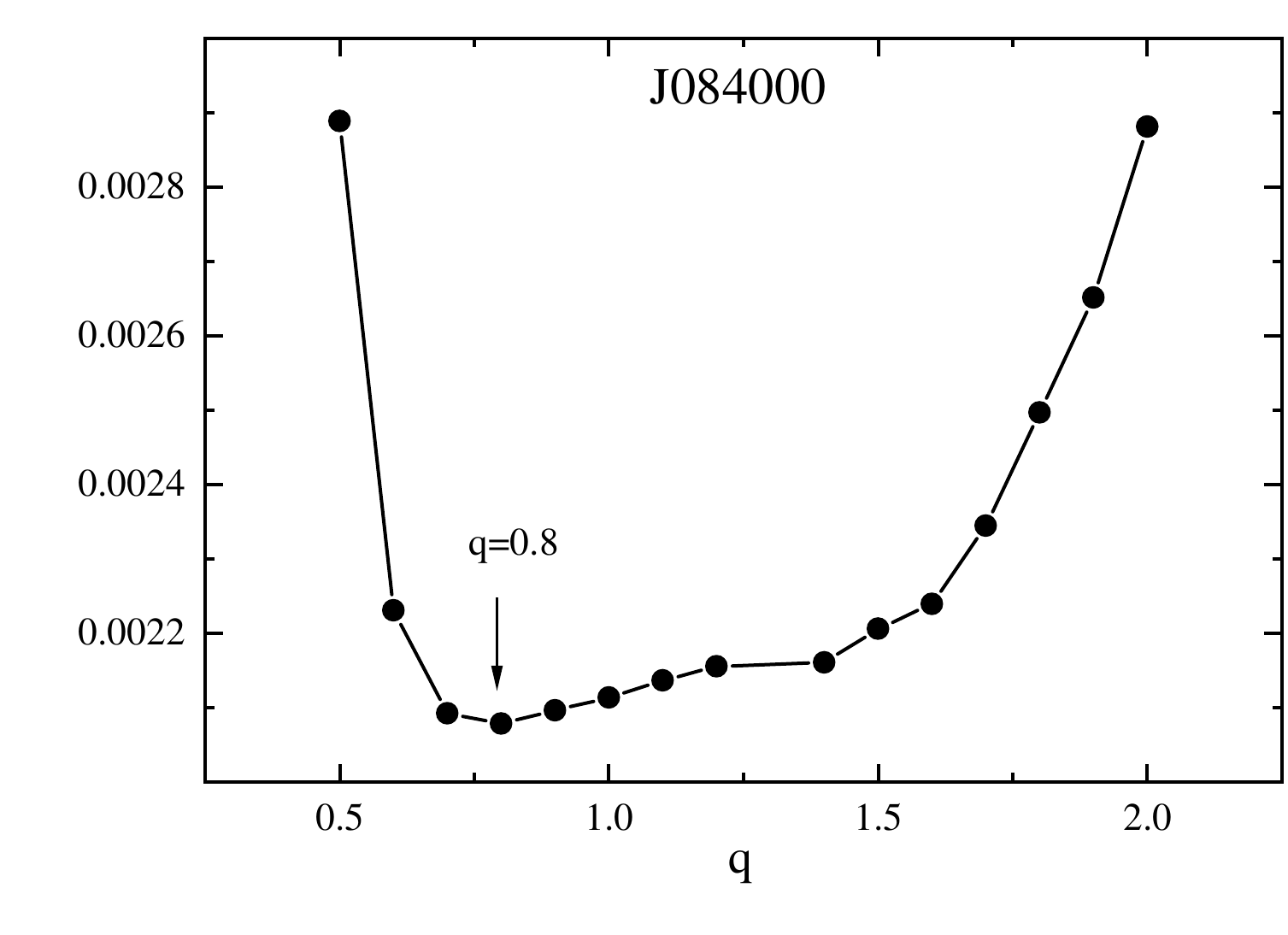}
 \caption{The relationships between mean residuals and mass ratio of the three targets.
\label{qsearch}}
\end{figure}
Then we used these new temperatures to model the light curves once again and obtained the final photometric solutions. The final converged solutions are presented in Tables \ref{tab:WD1}, \ref{tab:WD2}, \ref{tab:WD3}, while the synthetic and observed light curves are shown in Figure \ref{WD1}.
\par The light curves observed by ASAS-SN, TESS, ZTF, CRTS and SuperWASP were also analyzed by the W-D program. We downloaded TESS FFI cutouts \citep{2019AAS...23324510B} using the Lightkurve package\footnote{\url{https://lightkurve.github.io/lightkurve/}}  \citep{2018ascl.soft12013L} from the Mikulski Archive for Space Telescopes (MAST; DOI:\dataset[10.17909/3y7c-wa45] {http://dx.doi.org/10.17909/3y7c-wa45}). We created apertures and background masks using the thresholding method, and then used them to generate light curves. Subsequently, we converted the normalized flux to magnitude using the following formula:
\begin{equation}
	m = -2.5\lg E, 
\end{equation}
where m represents the magnitude and E is the normalized flux. We also cross-matched our targets with the Gaia website and set the radius as 42 arcsec. Our target stars are roughly at a magnitude of 14, while the brightest of the background stars have a magnitude of 18. Therefore, the impact of the third light is negligible. For the observational data of TESS, we used 200 bins per orbital period to create normal points and obtained quick solutions. If the light curves from different TESS sectors are similar, we analyzed them simultaneously, otherwise we analyzed them separately. For J034828, we integrated the data from Sectors 42, 43, and 44. Considering the 600 seconds observational cadence of the light curve, we accounted for phase smearing effects \citep{phasesmearingeffect} and chose NGA = 2 (the number of Gaussian quadrature abscissa in phase or time smearing simulation). Due to the differences in amplitude between Sector 71 (200s cadence) and the other three sectors, we analyzed it separately. For J073802, since the light curves from different sectors were similar, we put the data  together to obtain physical parameters. 
\par For the SuperWASP observational data, we used the same approach to combine the data into 200 points, then obtained the physical parameters. We used the parameters determined by our observations as initial values, set them as adjustable parameters and run the W-D program. The physical parameters are shown in Tables \ref{tab:WD1}, \ref{tab:WD2}, \ref{tab:WD3}. The theoretical light curves and observed light curves of TESS are shown in Figure \ref{WD1}, the light curves of other surveys are shown in Figure \ref{WD2}. The results from most of the other surveys are relatively consistent with ours. However, the solutions from surveys such as CRTS and SuperWASP show significant differences. This variation may be due to the photometric precision of CRTS and SuperWASP being too low for determining reliable physical parameters. Considering the observational accuracy and the multi-band nature of our data, we adopted the solutions obtained from our observations as the final results.

\vspace{-1cm}
\begin{deluxetable*}{cccccccc}
	\tablecaption{Photometric solutions of J034828.
		\label{tab:WD1}}
	\tablenum{6}
	\tablehead{\colhead{J034828} & \colhead{WHOT} & \colhead{TESS (600s)} & \colhead{TESS (200s)} & \colhead{ASAS-SN} & \colhead{ZTF} & \colhead{CRTS} & \colhead{SWASP} }
	\startdata
	$q$ & 2.438±0.021 & 2.389±0.011 & 2.506±0.007 & 2.463$\pm$0.015 & 2.406±0.027 & 2.463±0.015 & 2.395±0.231 \\
	$ T_1(K)$ & 5059±58 & 5097±51 & 5048±52 & 4952$\pm$96 & 5038±53 & 4979±53 & 4991±53 \\
	$ T_2(K)$ & 4831±81 & 4815±72 & 4847±75 & 4898$\pm$153 & 4852±76 & 4885±76 & 4880±76 \\
	$i(\degr)$ & 86.4±0.3 & 88.1±0.2 & 84.6±0.1 & 80.0$\pm$1.7 & 86.2±0.4 & 76.9±1.5 & 73.5±0.8 \\
	$\Omega_1=\Omega_2$ & 5.766±0.006 & 5.684±0.015 & 5.900±0.009 & 5.747$\pm$0.05 & 5.722±0.034 & 5.779±0.043 & 5.795±0.313 \\
	$( L_2/L_1)_R$ & 1.795±0.005 & — & — & — & — & — & — \\
	$( L_2/L_1)_I$ & 1.869±0.004 & — & — & — & — & — & — \\
	$( L_2/L_1)_V$ & — & — & — & 2.081$\pm$0.046 & — & 2.000±0.053 & — \\
	$( L_2/L_1)_r$ & — & — & — & — & 1.693±0.008 & — & — \\
	$( L_2/L_1)_i$ & — & — & — & — & 1.821±0.006 & — & — \\
	$( L_2/L_1)_{T}$  & — & 1.712±0.002 & 1.938±0.003 & — & — & — & — \\
	$( L_2/L_1)_{SWASP}$ & — & — & — & — & — & — & 1.956±0.071 \\
	$ r_1$ & 0.316±0.001 & 0.319±0.001 & 0.309±0.001 & 0.321$\pm$0.006 & 0.317±0.001 & 0.317±0.005 & 0.307±0.008 \\
	$ r_2$ & 0.470±0.001 & 0.470±0.001 & 0.468±0.001 & 0.477$\pm$0.004 & 0.469±0.005 & 0.473±0.005 & 0.459±0.042 \\
	$f$ & 15.5\% & 18.0\% & 8.8\% & 23.6\% & 15.6\% & 19\% & 4.4\% \\
	spot & star 1 & star 1 & star 1 & — & — & — & — \\
	$\theta(\degr)$ & 79±28 & 36±5 & 34±2 & — & — & — & — \\
	$\lambda(\degr)$ & 109±2 & 113±4 & 71±4 & — & — & — & — \\
	$r_s(\degr)$ & 25±5 & 25±5 & 26±1 & — & — & — & — \\
	$T_s$ & 0.862±0.006 & 0.878±0.006 & 0.879±0.006 & — & — & — & — \\
	\enddata
	\tablecomments{1. The TESS (600s) corresponds to the data of 600 seconds cadence, while the TESS (200s) corresponds to the data of 200 seconds cadence.\\
		2. The fill-out factor($f$) describes the contact degree of a contact binary, it can be calculated using $f = \frac{\Omega_{in}-\Omega}{{\Omega_{out}-\Omega}}$, where $\Omega_{in}$ and $\Omega_{out}$ are the potentials of the inner and outer Lagrangian equipotential surface, $\Omega$ is the actual potential.\\ 
		3. The spot parameters are latitude $\theta$, longitude $\lambda$, angular radius $r_s$ and the temperature factor $T_s$.}
	
\end{deluxetable*}

%表6

\begin{deluxetable*}{ccccccc}
	
	\tablecaption{Photometric solutions of J073802.
		\label{tab:WD2}}
	
	\tablenum{7}
	
	\tablehead{\colhead{J073802} & \colhead{NEXT} & \colhead{TESS} & \colhead{ASAS-SN} & \colhead{ZTF} & \colhead{CRTS} &\colhead{SWASP}  } 
	
	\startdata
	$q$ & 0.668±0.003 & 0.650±0.003 & 0.668±0.035 & 0.689±0.004 & 0.668±0.025 & 0.668$\pm$0.040 \\
	$T_1$(K) & 5651±56 & 5663±56 & 5681±58 & 5644±57 & 5711±62 & 5707$\pm$66 \\
	$T_2$(K) & 5712±94 & 5738±93 & 5668±95 & 5718±95 & 5624±108 & 5605$\pm$118 \\
	$i(\degr)$ & 89.7±0.1 & 89.0±0.1 & 85.1±0.9 & 89.1±0.3 & 88.1±2.4 & 89.7$\pm$3.1 \\
	$\Omega_1=\Omega_2$ & 3.147±0.005 & 3.117±0.001 & 3.181±0.062 & 3.196±0.007 & 3.124±0.012 & 3.038$\pm$0.018 \\
	$(L_2/L_1)_g$ & 0.738±0.002 & — & — & 0.761±0.004 & — & — \\
	$(L_2/L_1)_r$ & 0.726±0.002 & — & — & 0.752±0.003 & — & — \\
	$(L_2/L_1)_i$ & 0.721±0.002 & — & — & 0.746±0.003 & — & —\\
	$(L_2/L_1)_V$ & — & — & 0.681±0.021 & — & 0.645±0.010 & — \\
	$(L_2/L_1)_{T}$ & — & 0.709±0.001 & — & — & — & — \\
	$(L_2/L_1)_{SWASP}$ & — & — & — & — & — & 0.646$\pm$0.001 \\
	$r_1$ & 0.424±0.001 & 0.426±0.001 & 0.417±0.032 & 0.419±0.001 & 0.429±0.002 & 0.448$\pm$0.004 \\
	$r_2$ & 0.353±0.099 & 0.351±0.002 & 0.346±0.023 & 0.354±0.003 & 0.358±0.003 & 0.448$\pm$0.004 \\
	$f$ & 10.0\%±1.3\% & 9.8\%±0.3\% & 5.0\%±15.8\% & 7.1\%±1.8\% & 19.7\%±3.0\% & 38.8\%$\pm$4.7\% \\
	spot & star 1 & star 1 & — & — & — & — \\
	$\theta(\degr)$ & 21$\pm$1 & 11±1 & — & — & — & — \\
	$\lambda(\degr)$ & 54±2 & 40±3 & — & — & — & — \\
	$r_s(\degr)$ & 31±1 & 32±1 & — & — & — & — \\
	$T_s$  & 0.879±0.006 & 0.879±0.007 & — & — & — & — \\
	\enddata
	
\end{deluxetable*}

%表7
\begin{deluxetable*}{cccccc}
	\tablecaption{Photometric solutions of J084000.
		\label{tab:WD3}}
	\tablenum{8}
	
	\tablehead{\colhead{J084000} & \colhead{XL60} & \colhead{ASAS-SN} & \colhead{ZTF} & \colhead{CRTS} & \colhead{SWASP}  } 
	
	\startdata
	$q$ & 0.780±0.005 & 0.780±0.006 & 0.780±0.005 & 0.780±0.005 & 0.815±0.009 \\
	$T_1$(K) & 4910±71 & 4877±70 & 4885±70 & 4901±70 & 4941±70 \\
	$T_2$(K) & 4798±113 & 4747±115 & 4833±114 & 4812±113 & 4781±112 \\
	$i(\degr)$ & 85.3±0.1 & 82.2±0.4 & 84.7±0.2 & 82.2±0.9 & 84.8±0.3 \\
	$\Omega_1=\Omega_2$ & 3.320±0.012 & 3.364±0.008 & 3.351±0.004 & 3.282±0.023 & 3.424±0.015 \\
	$(L_2/L_1)_R$& 0.712±0.001 & — & — & — & — \\
	$(L_2/L_1)_I$ & 0.725±0.009 & — & — & — & — \\
	$(L_2/L_1)_V$ & — & 0.671±0.002 & — & 0.716±0.002 & — \\
	$(L_2/L_1)_g$ & — & — & 0.738±0.001 & — & — \\
	$(L_2/L_1)_r$ & — & — & 0.754±0.001 & — & — \\
	$(L_2/L_1)_i$ & — & — & 0.761±0.001 & — & — \\
	$(L_2/L_1)_{SWASP}$ & — & — & — & — & 0.677±0.001 \\
	$r_1$ & 0.414±0.001 & 0.406±0.001 & 0.408±0.001 & 0.422±0.005 & 0.402±0.001 \\
	$r_2$ & 0.371±0.003 & 0.362±0.001 & 0.355±0.001 & 0.379±0.005 & 0.366±0.001 \\
	$f$ & 14.4\%±2.6\% & 4.5\%±1.8\% & 7.4\%±1.8\% & 23.1\%±5.1\% & 4.0\%±3.2\% \\
	spot & star 1 & star 1 & — & — & — \\
	$\theta(\degr)$ & 57±2 & 17±3 & — & — & — \\
	$\lambda(\degr)$ & 119±3 & 170±6 & — & — & — \\
	$r_s(\degr)$ & 22±1 & 34±3 & — & — & — \\
	$T_s$ & 0.879±0.004 & 0.879±0.029 & — & — & — \\
	\enddata
	
\end{deluxetable*}

\begin{figure}
  \epsscale{0.36}
     \plotone{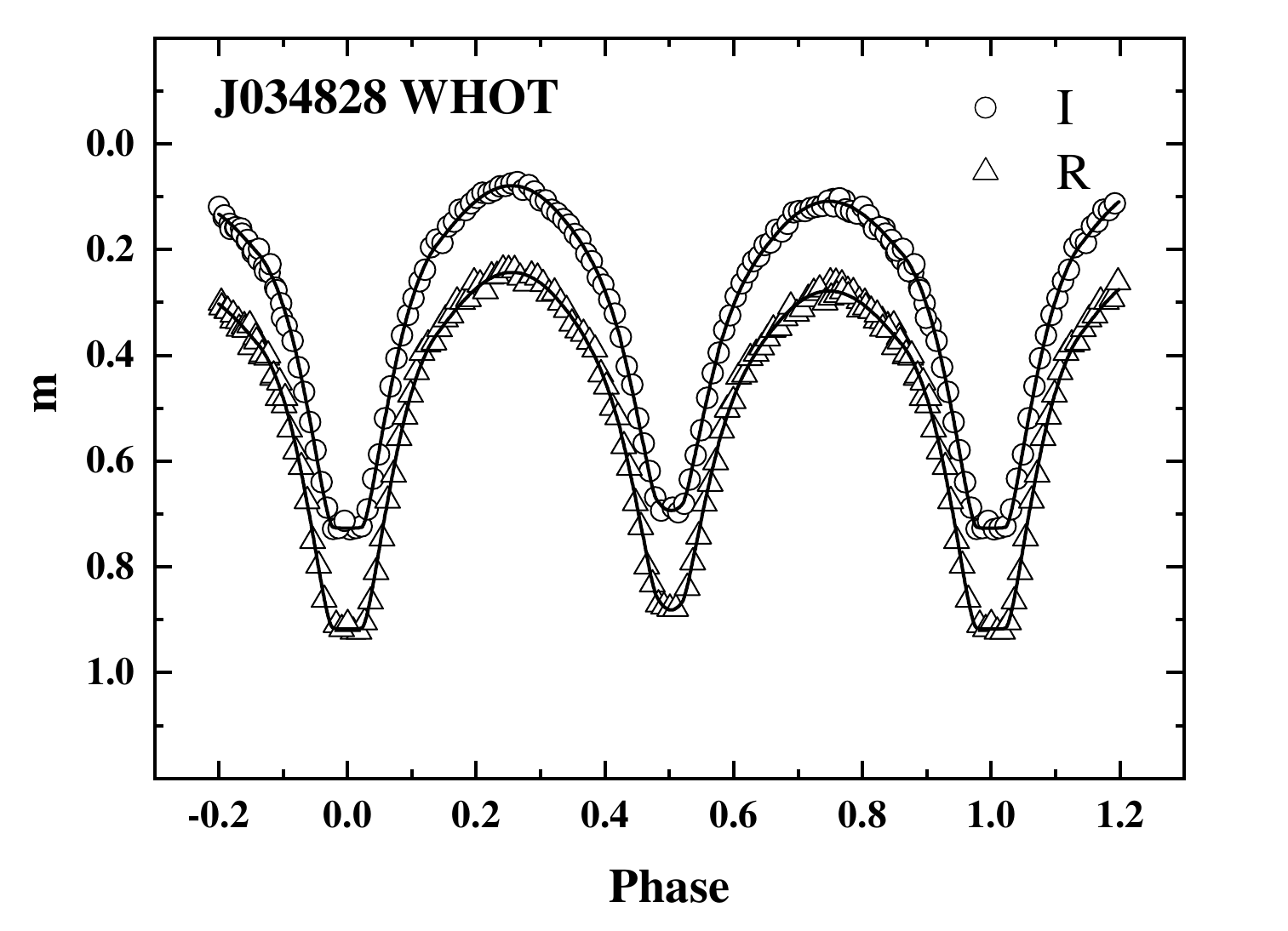}
	 \plotone{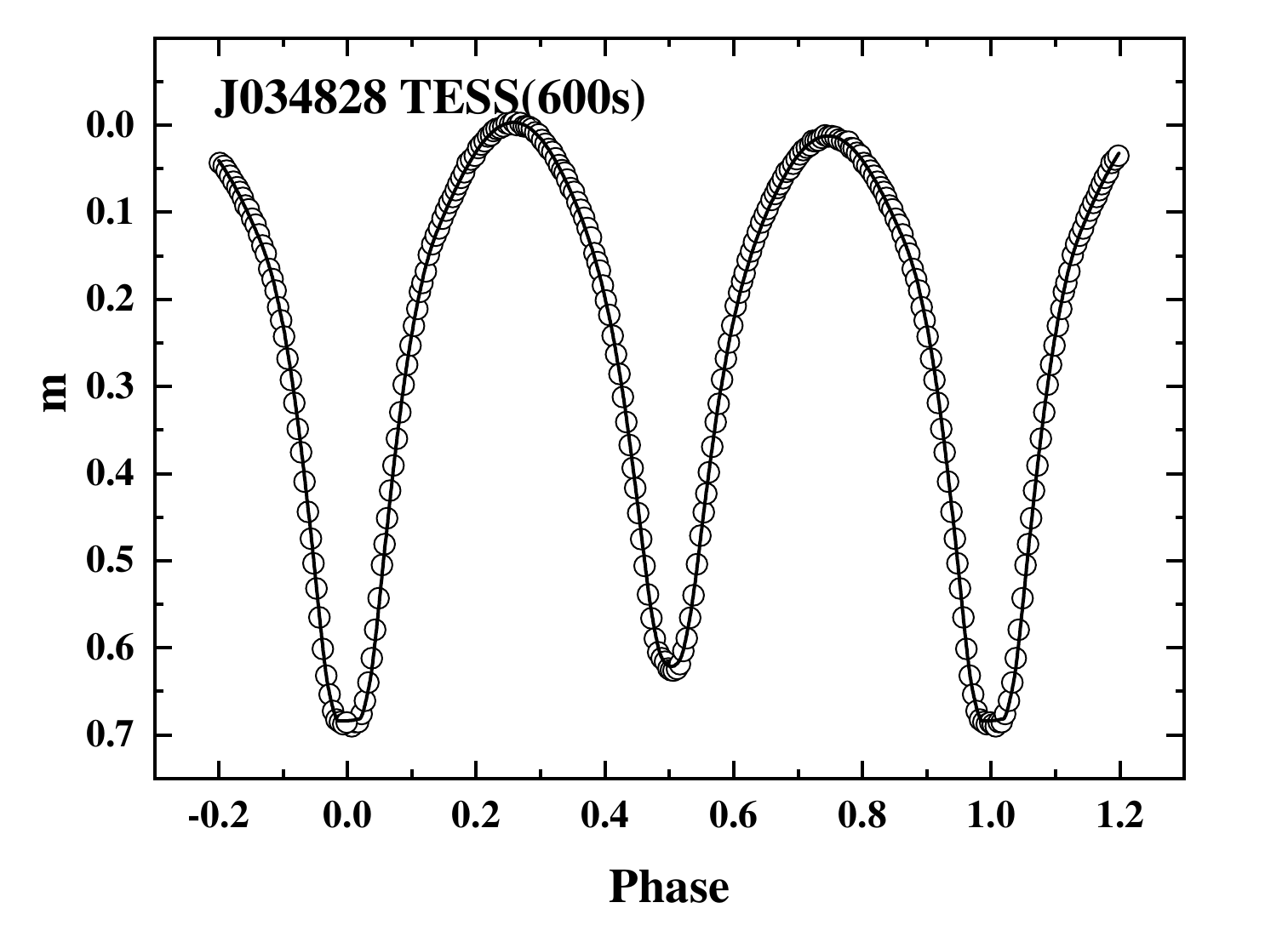}
	 \plotone{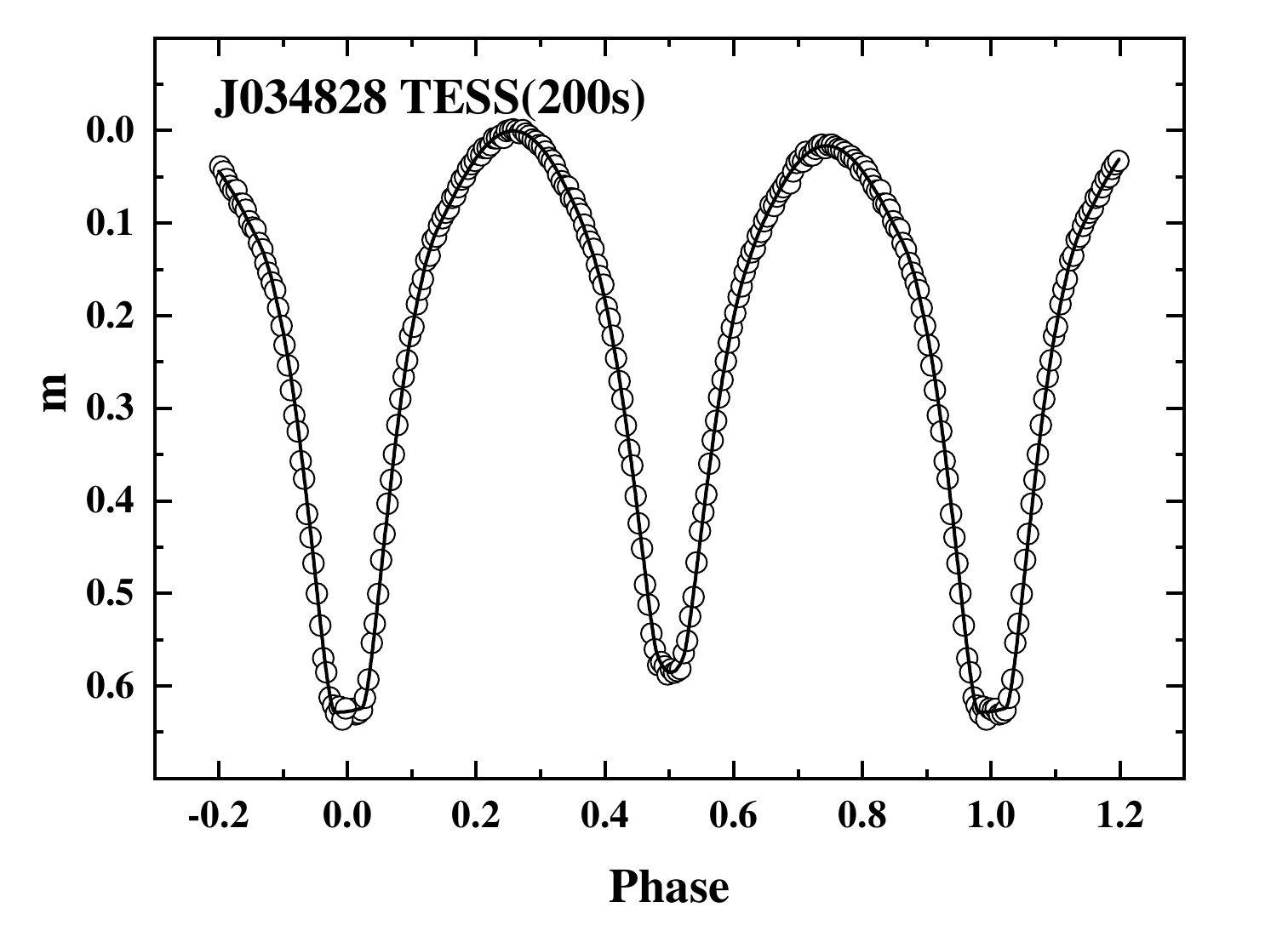}\\
     \plotone{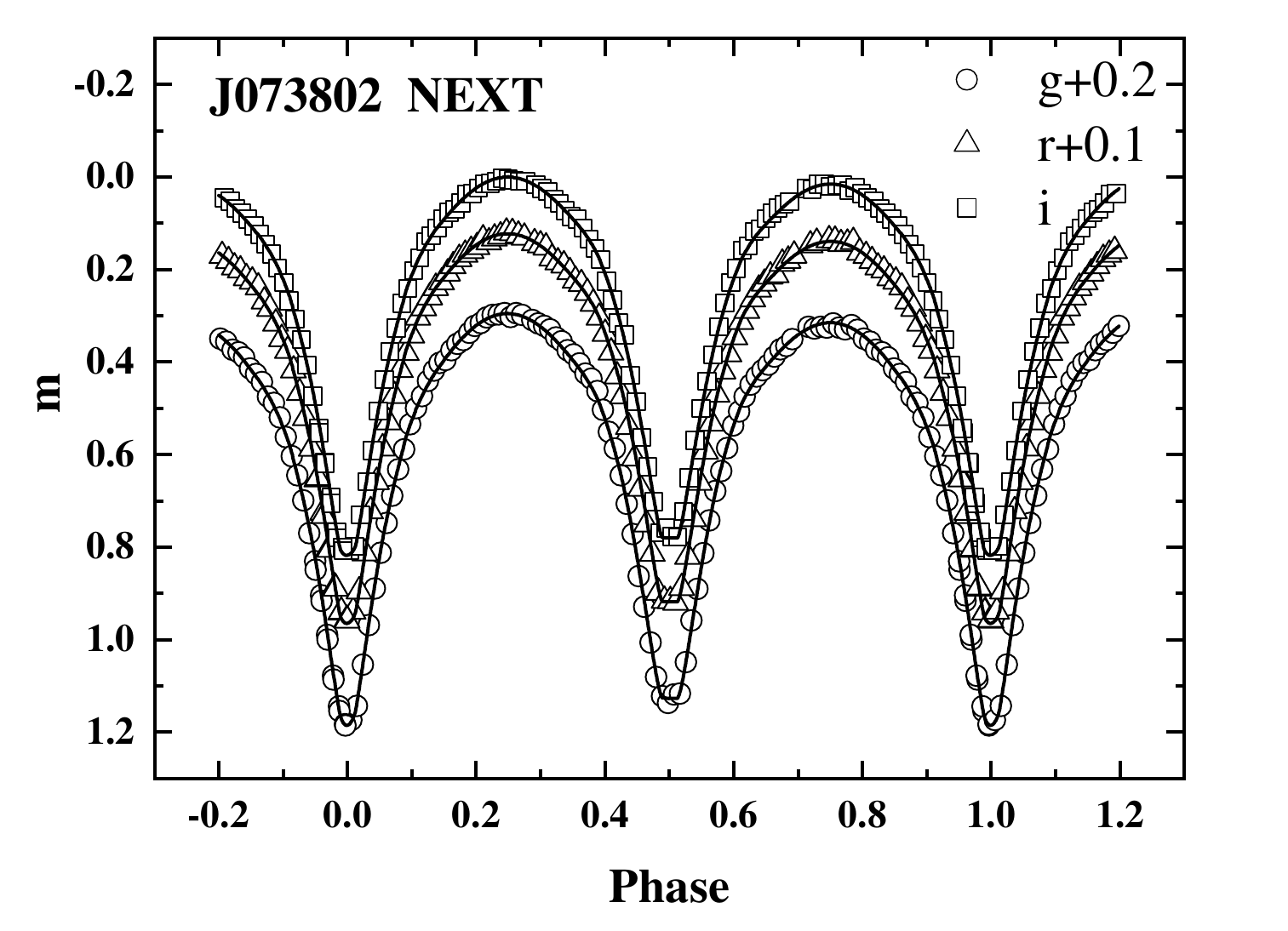}
	 \plotone{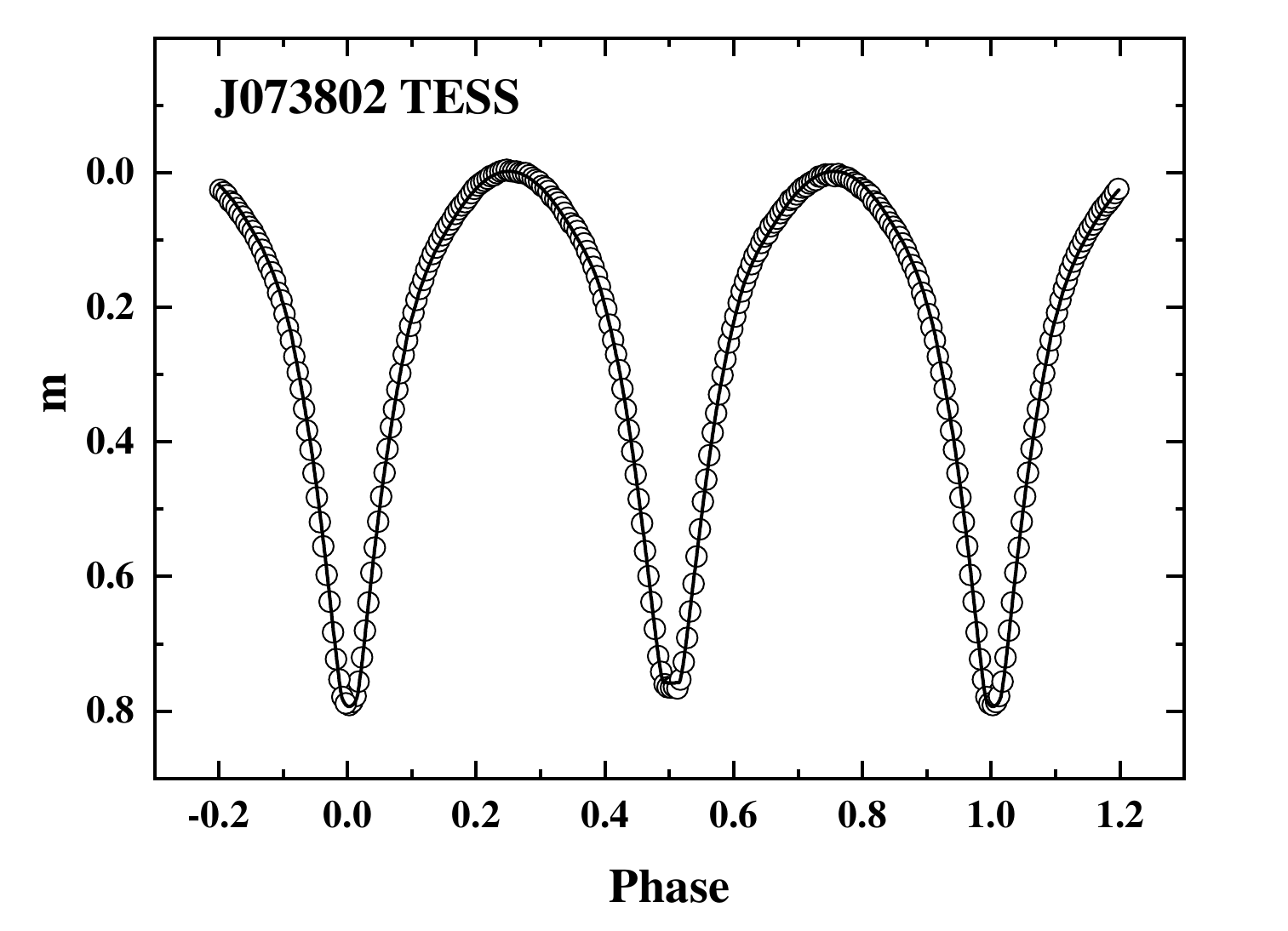}
     \plotone{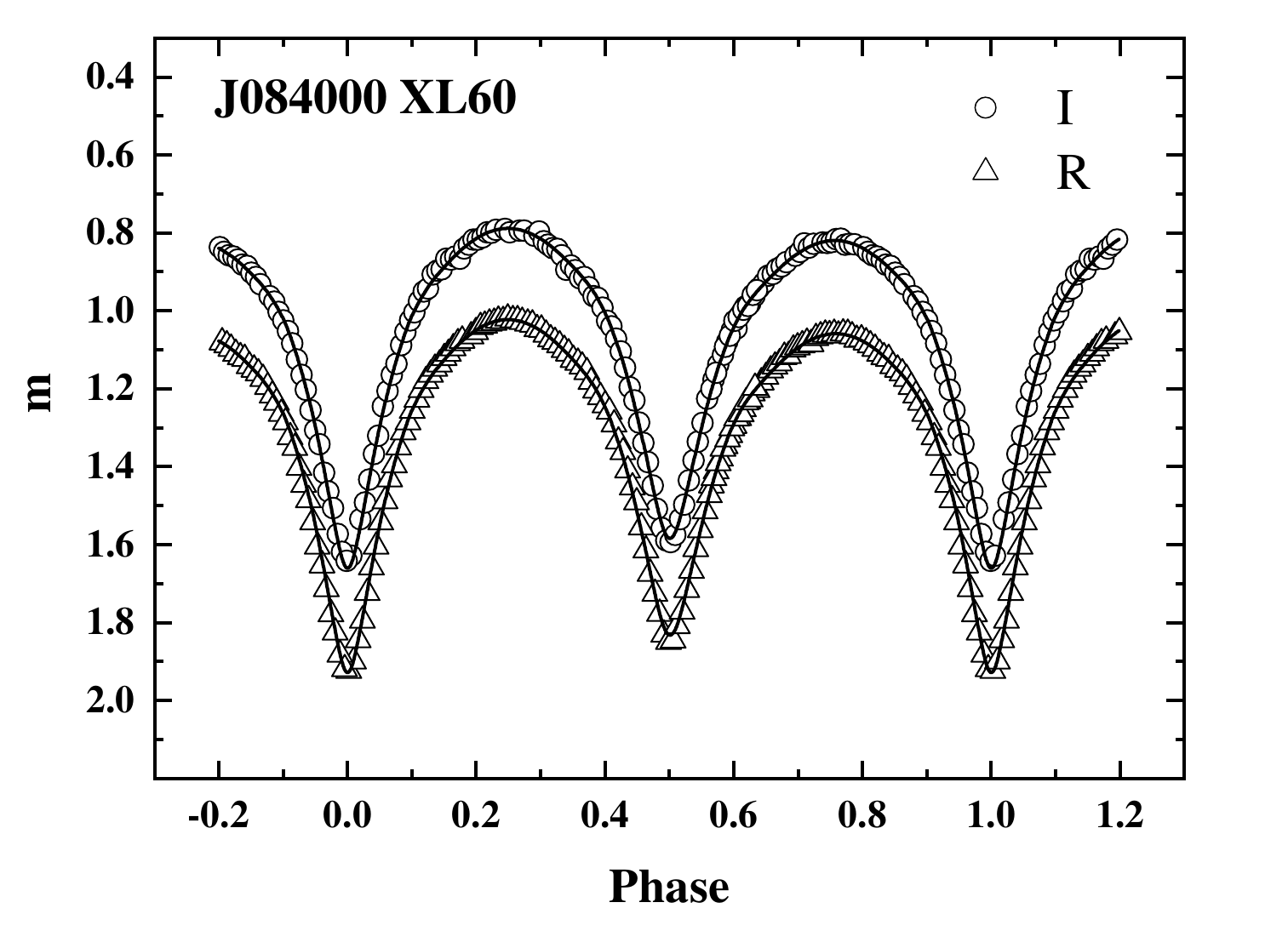}
	
	\caption{The theoretical curves and observed curves of our observations and TESS.
	\label{WD1}}
\end{figure}

\begin{figure}
   \epsscale{0.36}
	\plotone{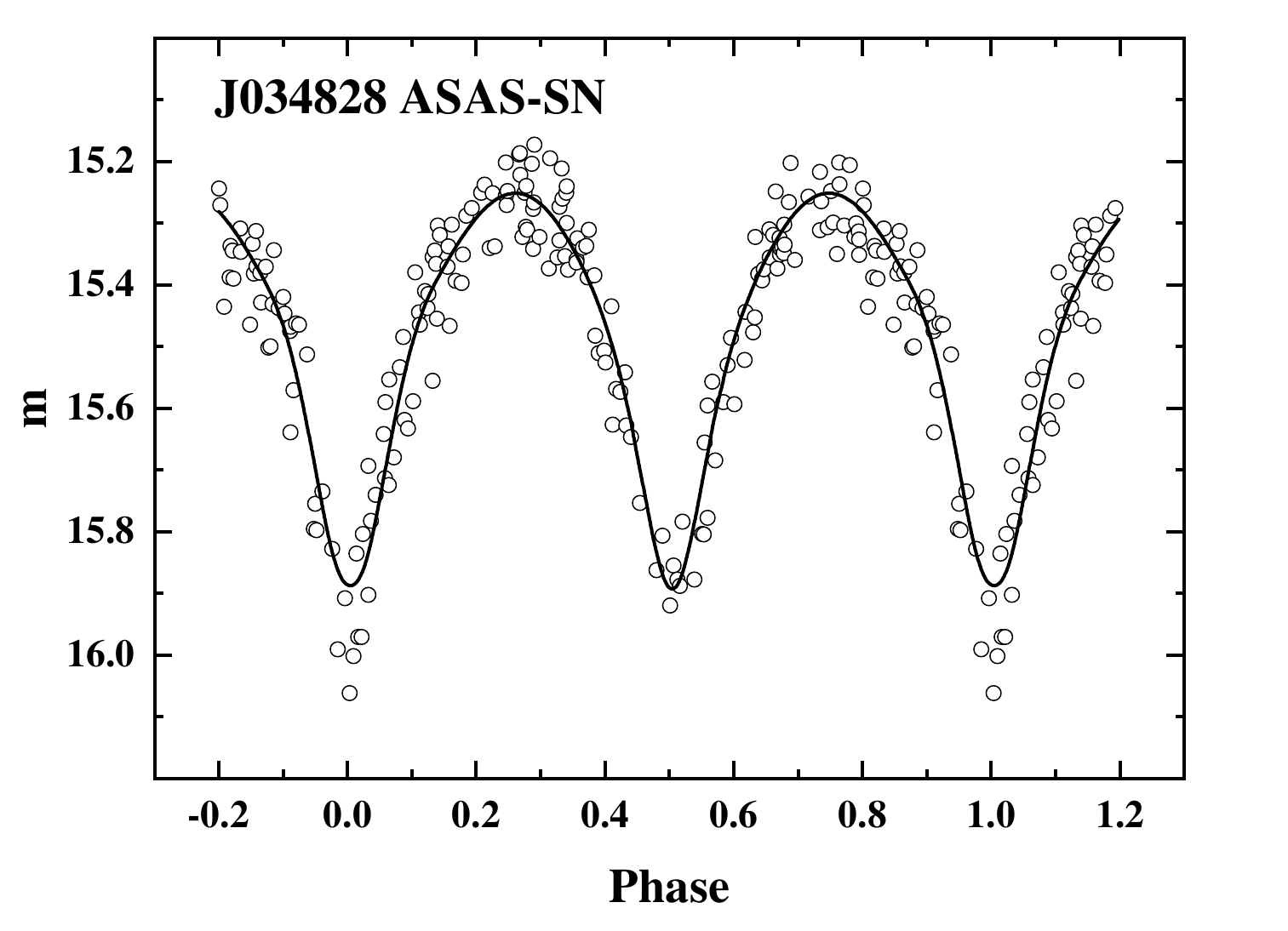}
	\plotone{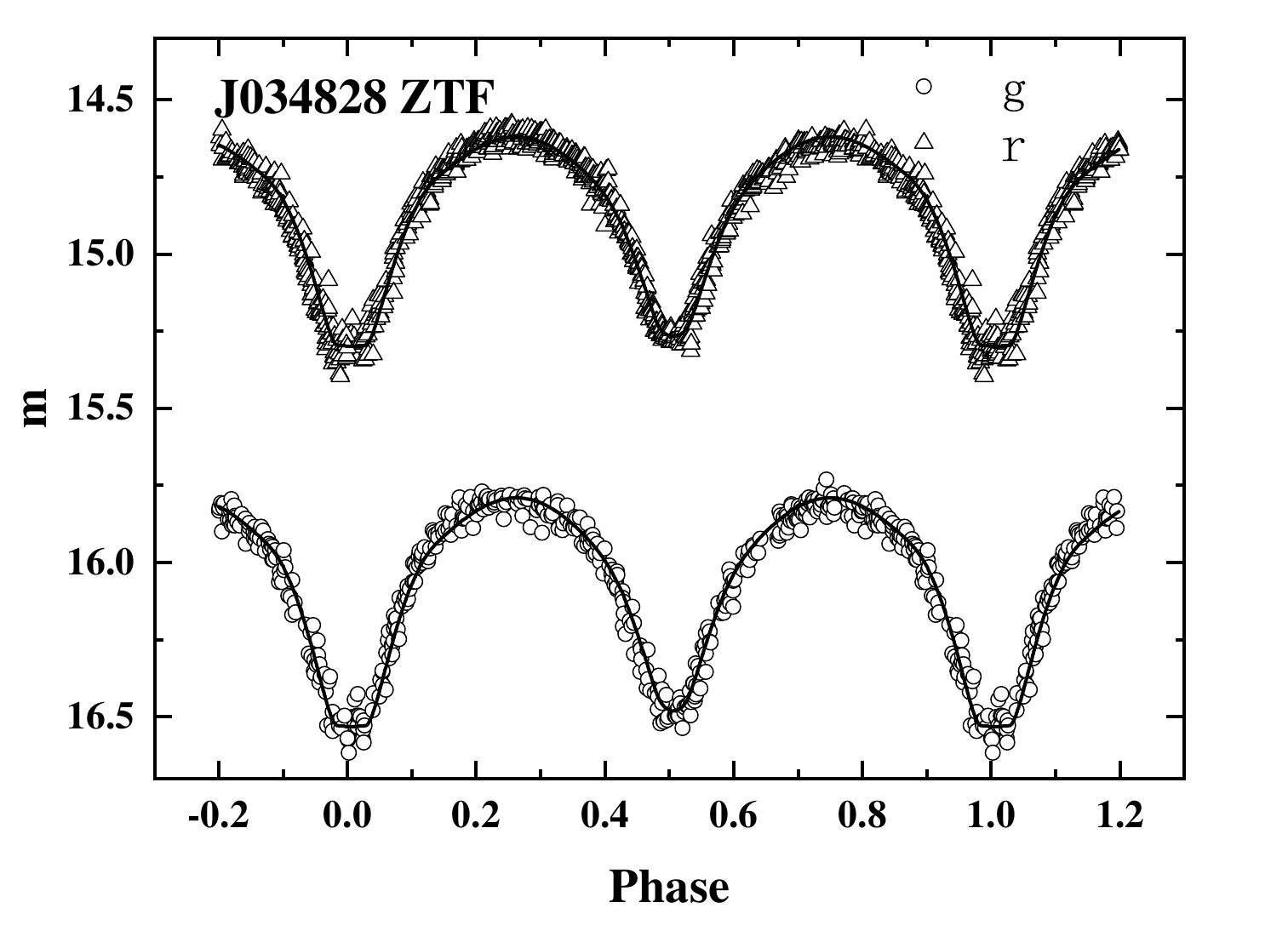}
    \plotone{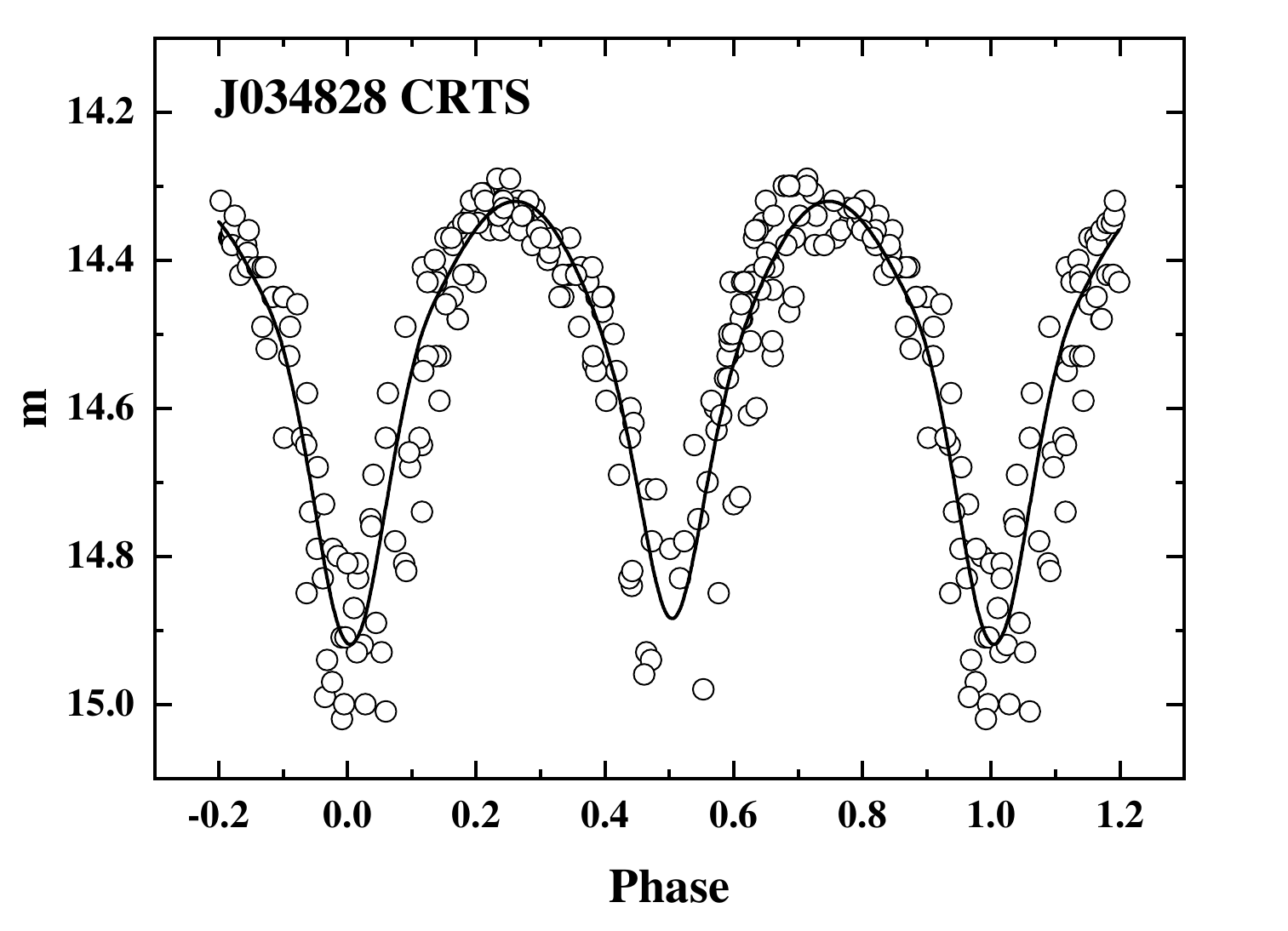}\\
	\plotone{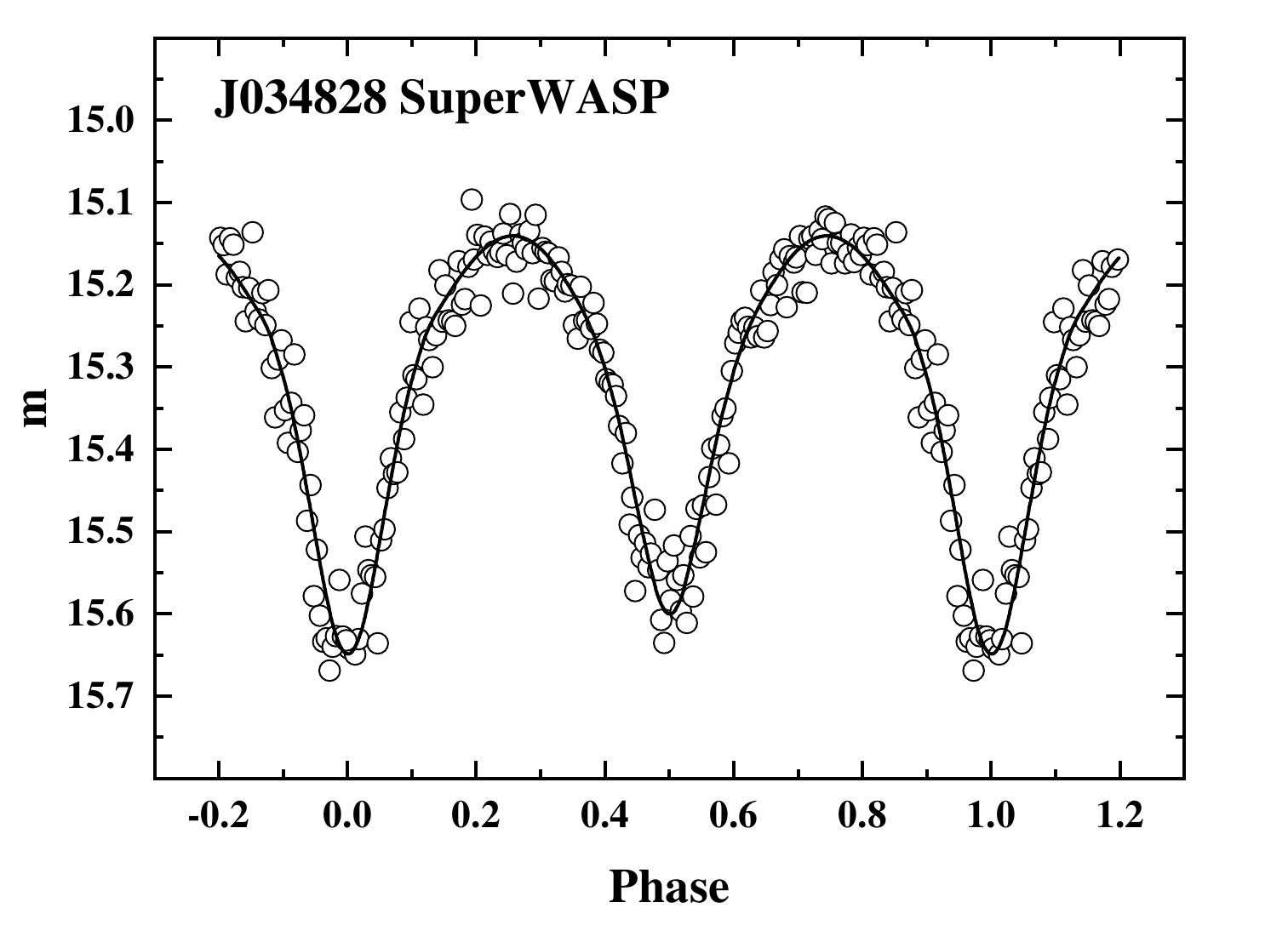}
    \plotone{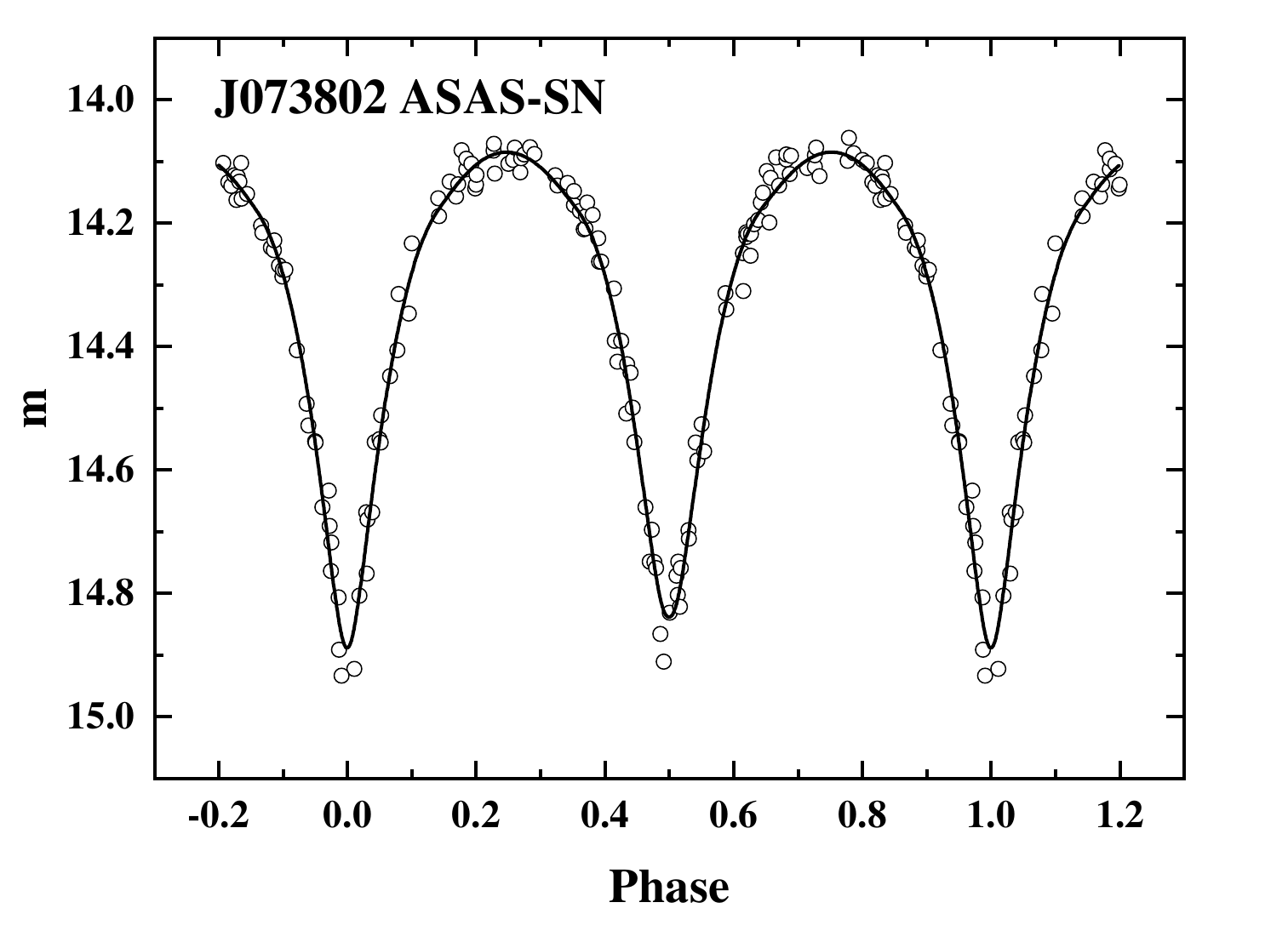}
    \plotone{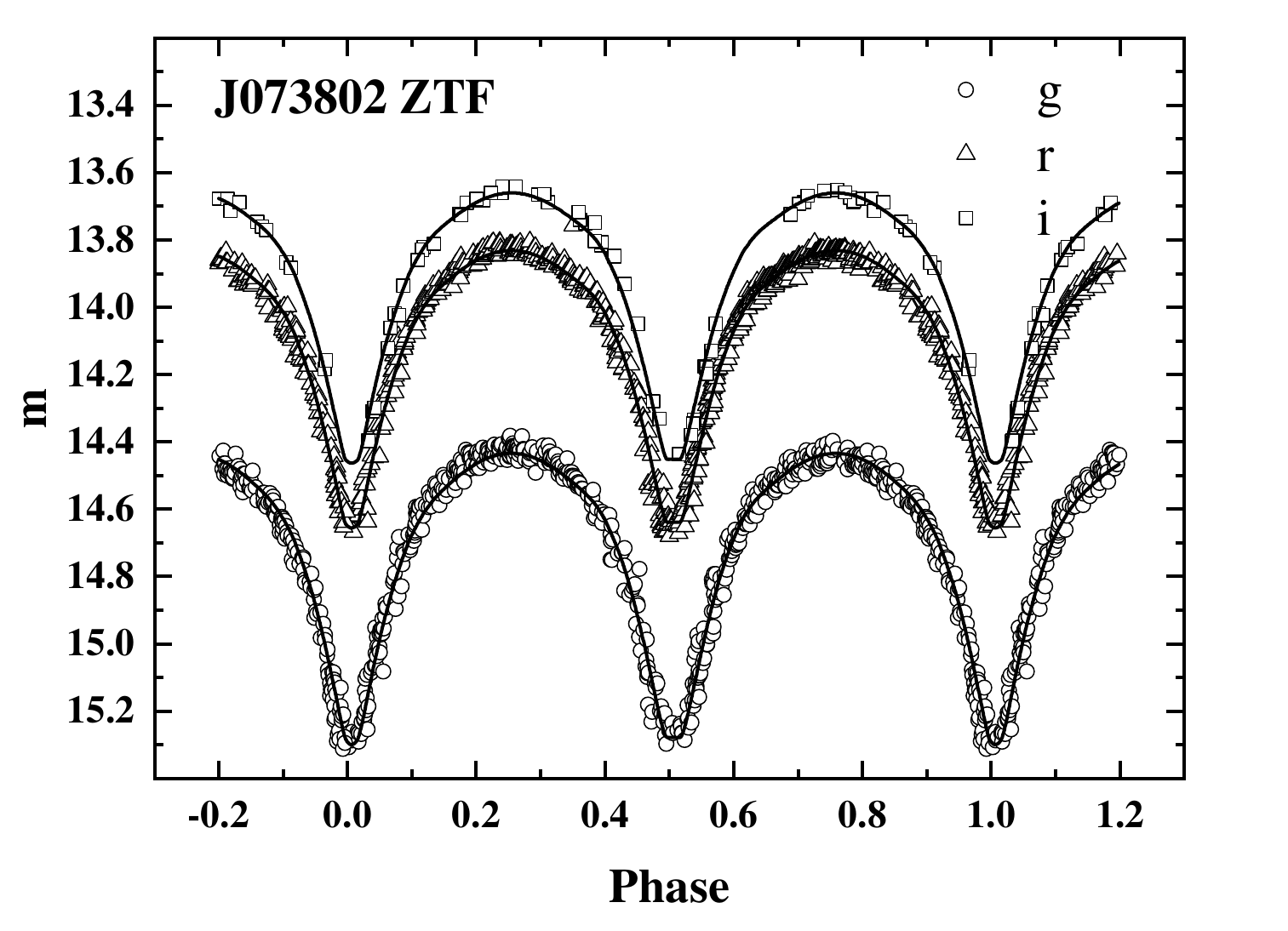}\\
	\plotone{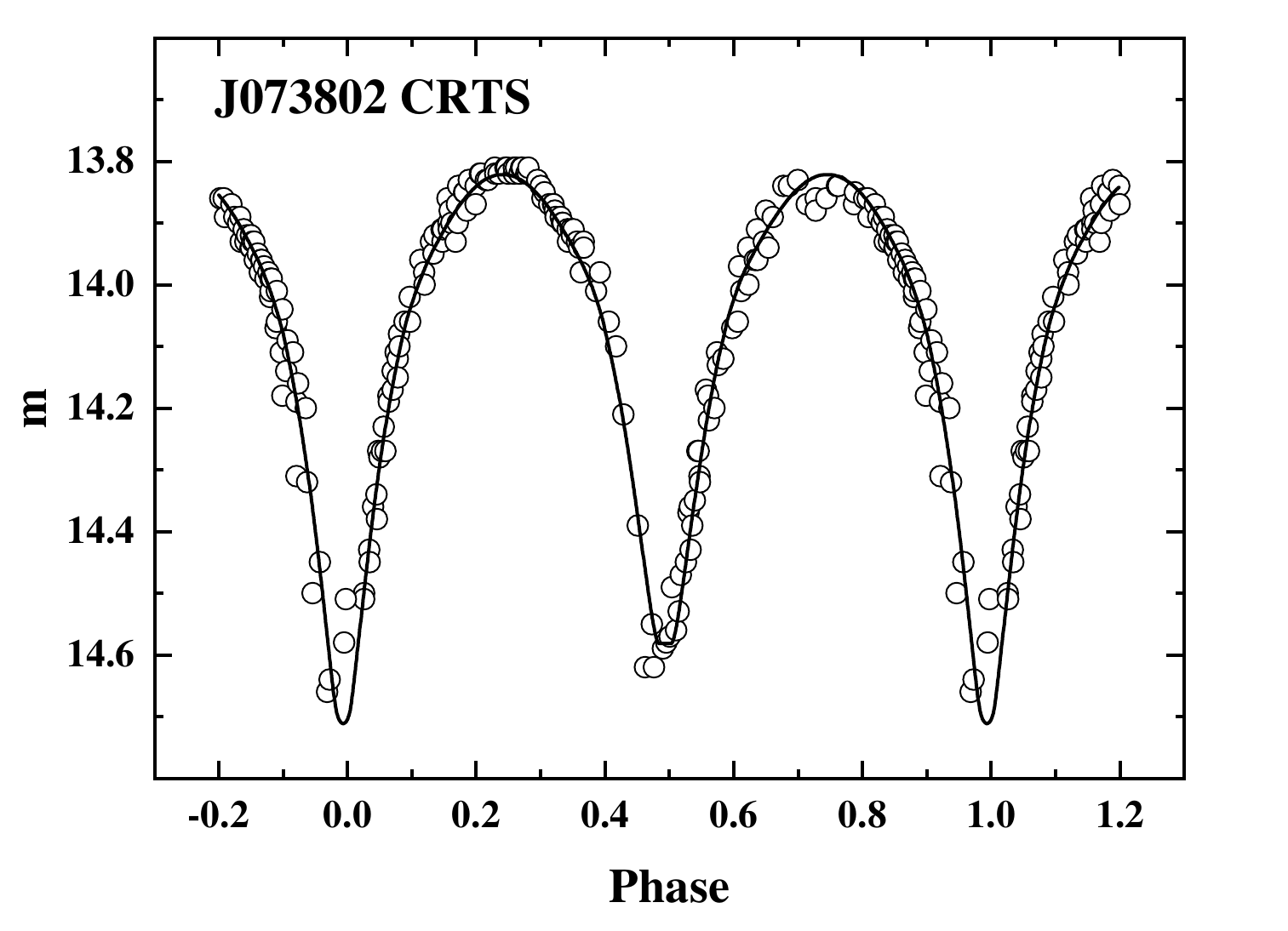}
	\plotone{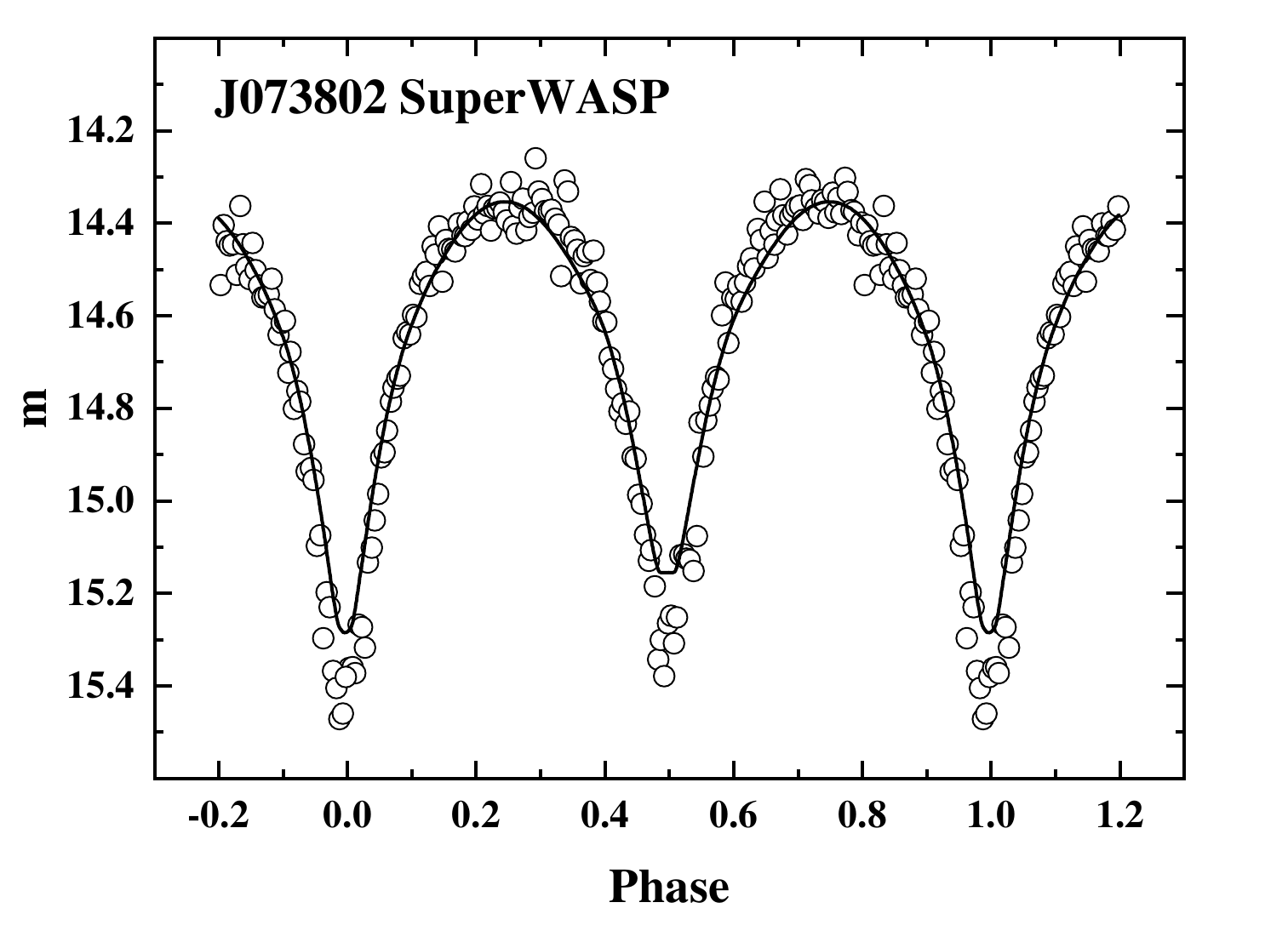}
	\plotone{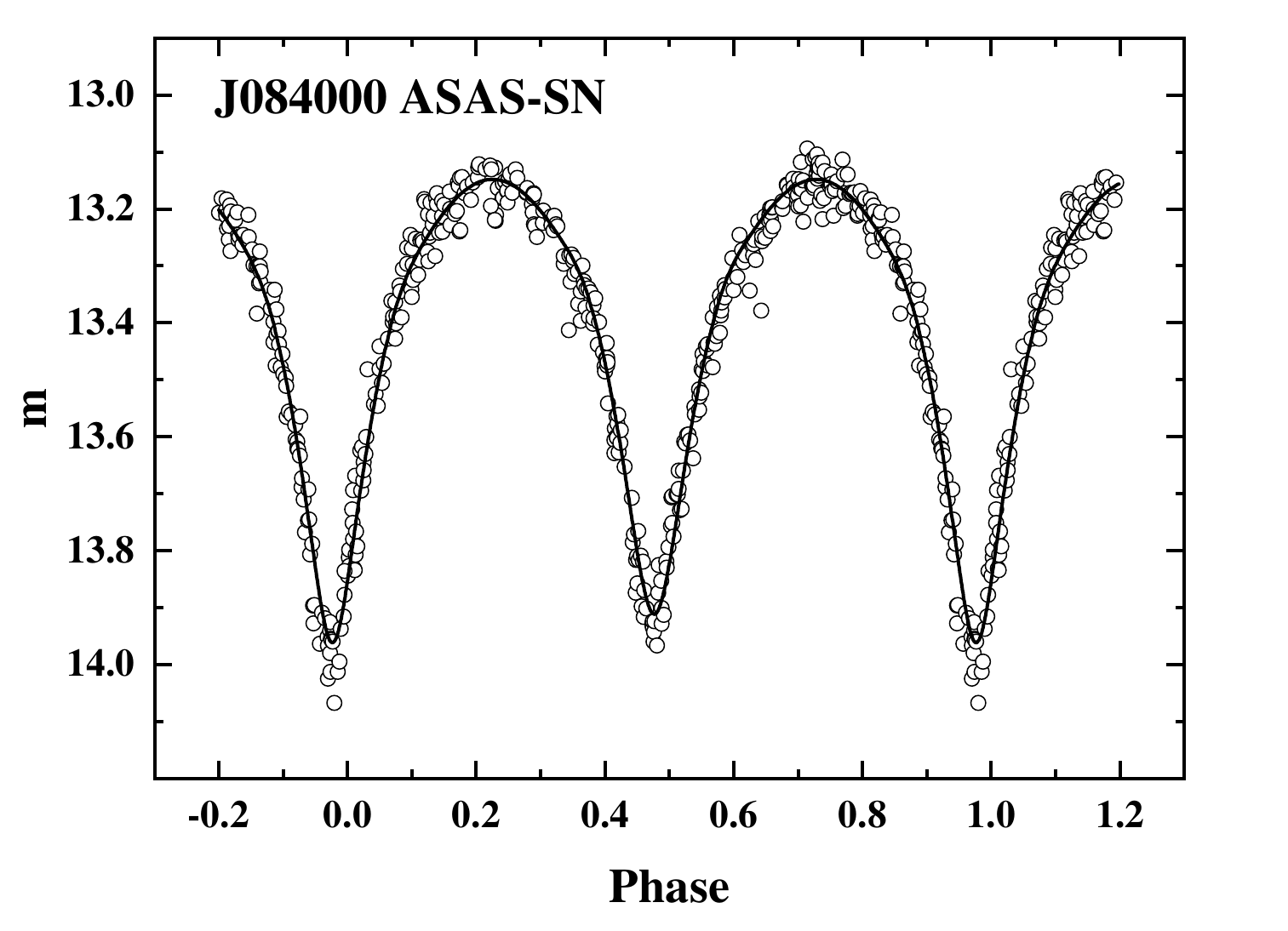}\\
	\plotone{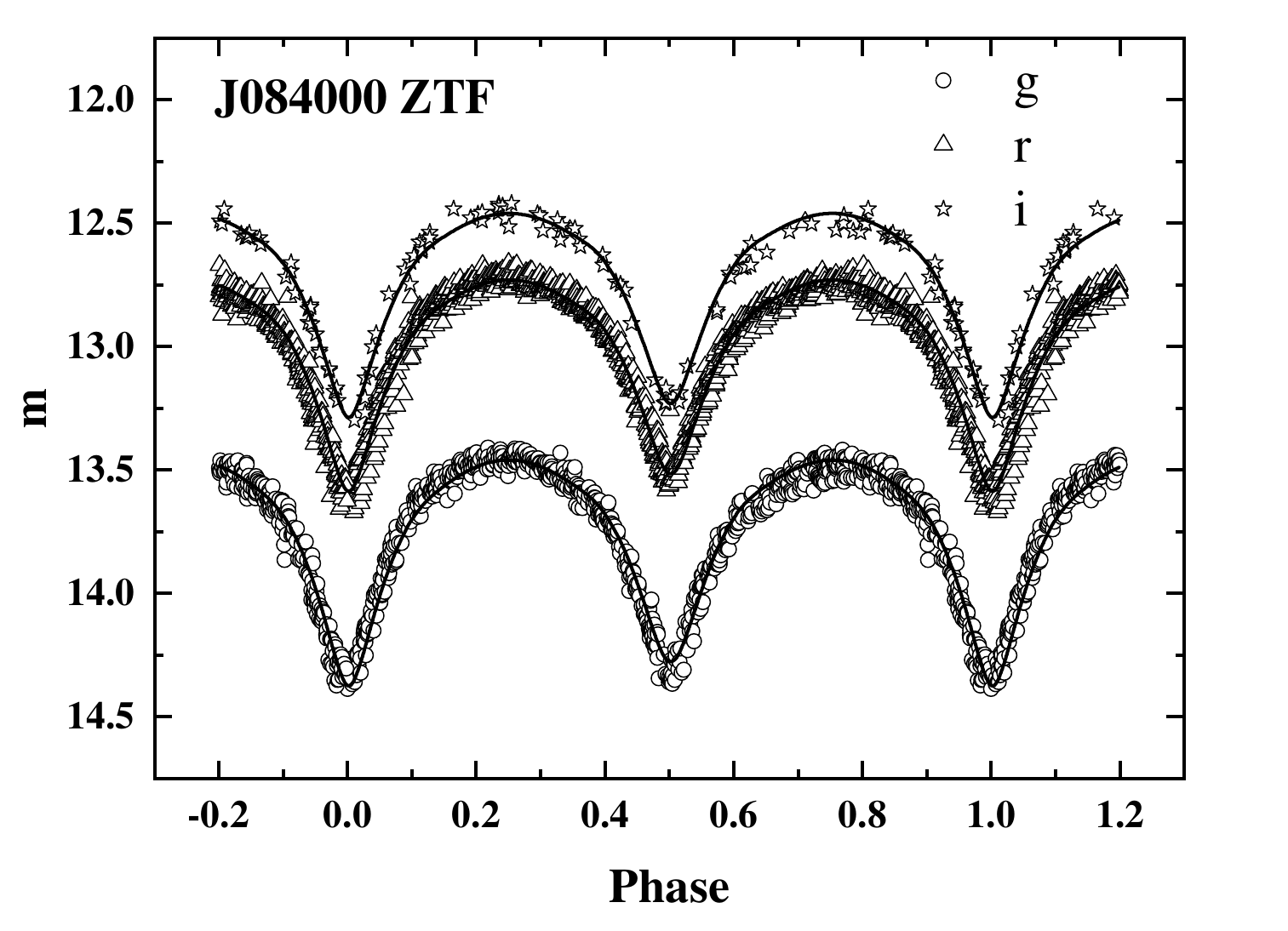}
	\plotone{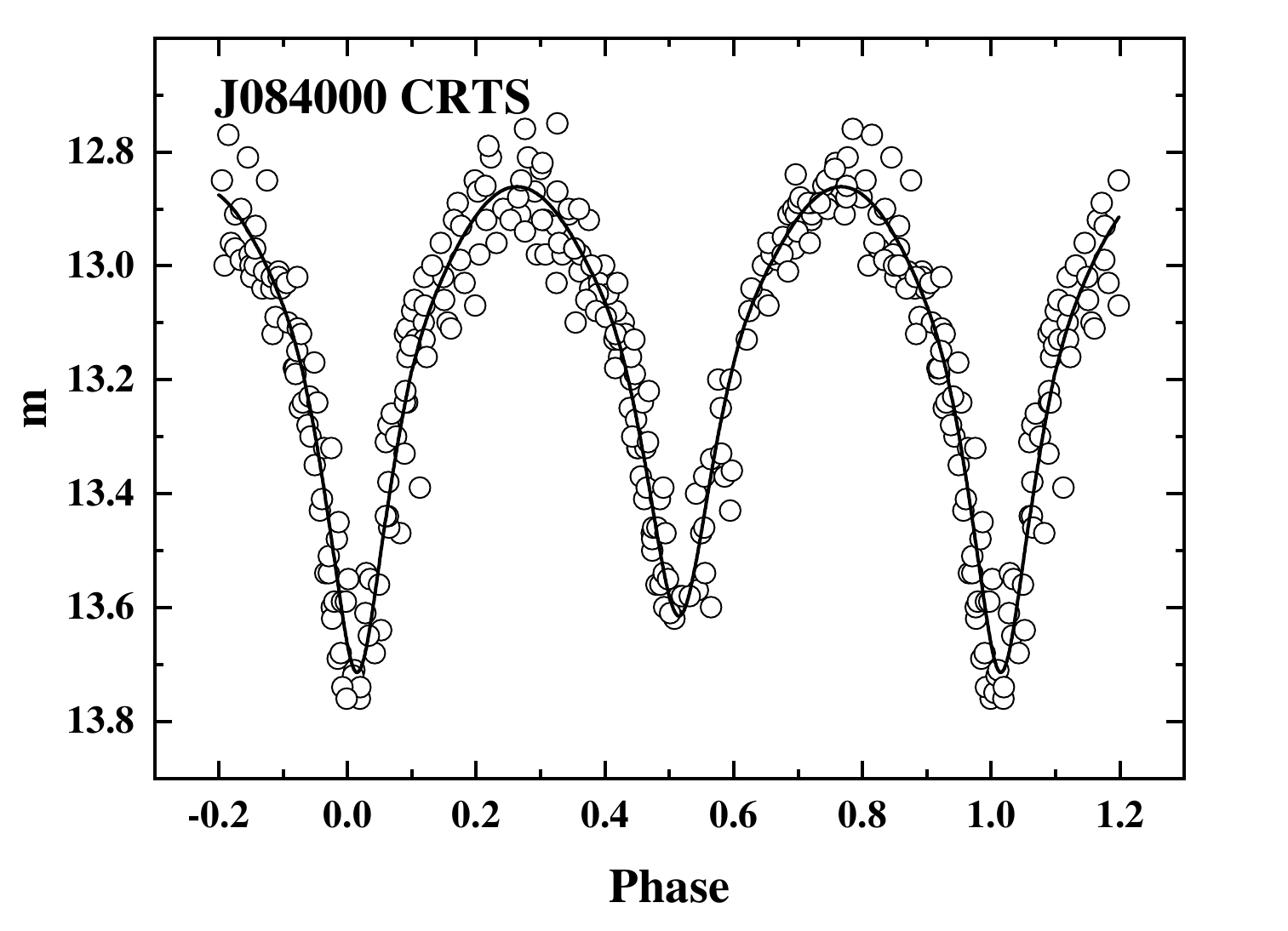}
    \plotone{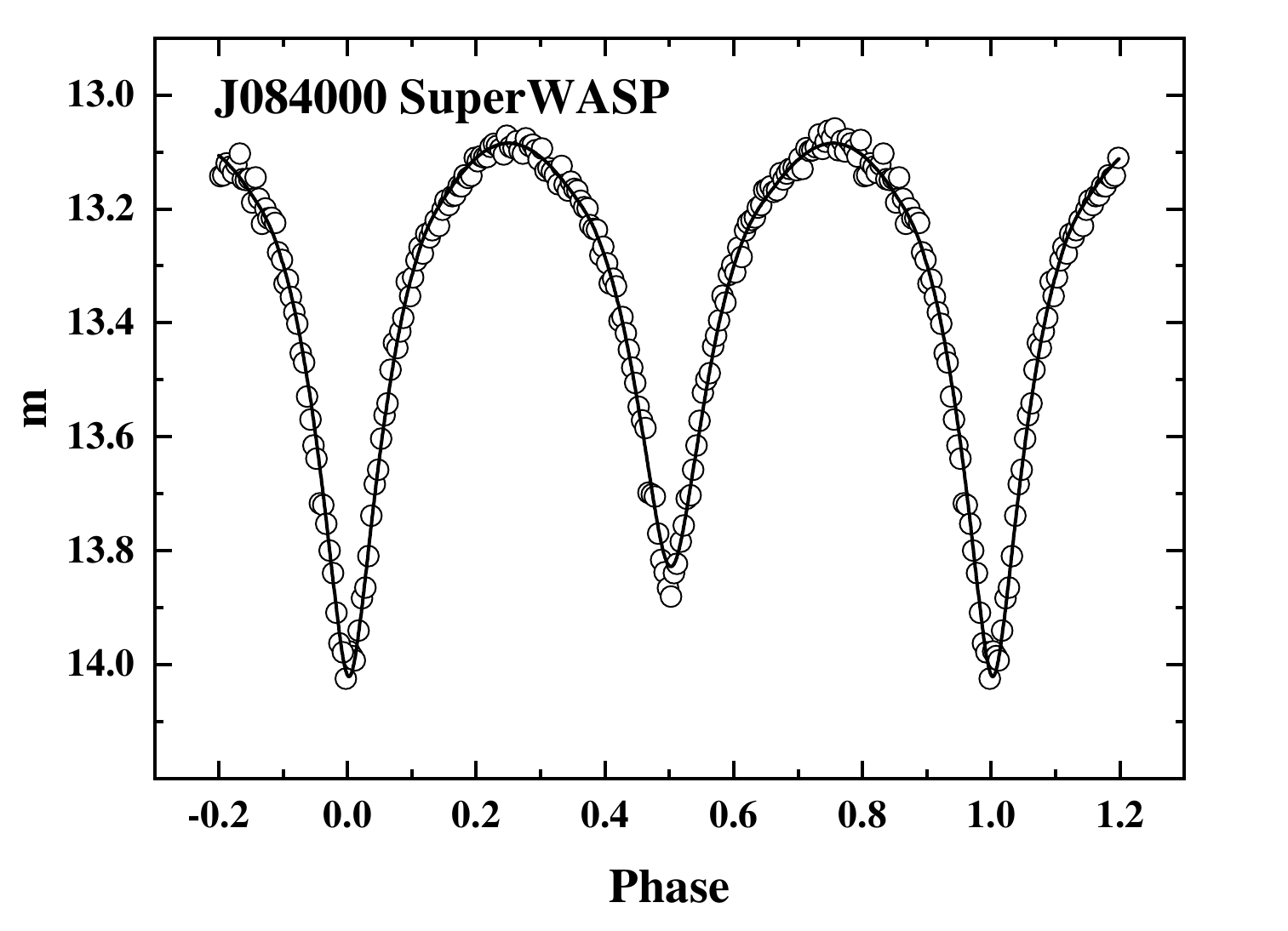}
	\caption{The theoretical curves and observed curves of other observations.
	\label{WD2}}
\end{figure}

\section{Orbital Period Analysis}
The variation of orbital period caused by angular momentum loss (AML) or mass transfer between the two components is common in contact binaries. Therefore, conducting an orbital period analysis is essential for studying the evolution of these systems. To accurately correct the orbital periods of three targets and obtain their period variations, we collected as many eclipsing times as possible from our observations, as well as data from ASAS-SN, TESS, SuperWASP, CRTS, ZTF, and the literatures for O-C analysis. Accounting for the continuous observations of TESS and SuperWASP light curves, we employed the K-W method \citep{KW} to directly calculate the times of minima. For ASAS-SN, CRTS, and ZTF, given that the data of them are discrete, the period shift method suggested by \citet{2020AJLi,Gaia,massratiolimit1} was used to translate observational data in to one period, then we used K-W method to calculate the minima. Based on the following equation, we divided the light curves into segments and synchronized them to one period,
\begin{equation}
\text{HJD/MJD}=\text{HJD}_0/\text{MJD}_0+P\times E, 
\end{equation}
where HJD/MJD represents the observing time, $\text{HJD}_0/\text{MJD}_0$ is the reference time, P is the period and E stands for the cycle number. All the calculated eclipsing times were converted to Barycentric Julian Date (BJD) using the online calculator\footnote{\url {https://astroutils.astronomy.osu.edu/time/hjd2bjd.html}} provided by \citet{BJD} and listed in Table \ref{tab:O-C1}.
The O-C values of three targets can be calculated using the following linear ephemeris:
\begin{equation}
	T= T_0+P\times E, 
\end{equation}
where $T$ represents the observed eclipsing times, $T_0$ is the initial primary minimum. The O–C values are listed in Table \ref{tab:O-C1}. We applied simple linear fitting to model their O–C diagrams and corrected the orbital period at first, the corrected values are displayed in Table \ref{tab:O-C2}. Then we found there is no parabolic tendency for J073802 and J084000, but an obvious parabolic tendency exists in J034828. Therefore, the following equation was used to model its O–C diagram,
\begin{equation}
	O-C = \Delta T_0+ \Delta P \times E + \frac{\beta}{2}\times E^2,
\end{equation}
where $\Delta T_0$ is the correction of initial primary minimum, $\Delta P$ is the correction of orbital period, and $\beta$ quantifies the rate of long term orbital period change.  We determined the orbital period of J034828 is long term increasing at the rate of $dP/dt = 1.73\pm0.05 \times10^{-7} day \ yr^{-1}$. We also fitted the O-C diagram of J034828 using a periodic sine function. The upward trend in the O-C diagram can be regarded as part of the cyclic variation caused by the light-time effect, and the new fitting curve is represented by the red line in Figure \ref{O-C}. The fitting equation is given by:
	\begin{equation}
		O-C = 0.0031(\pm 0.0002) + 0.0089(\pm 0.0004) \times \sin[0.00012(\pm0.00002) \times E - 0.10(\pm0.03)].
\end{equation}
Based on this equation, a cyclic variation with an amplitude of 0.0089 days and period of 34.81 years were obtained. The O–C diagrams of the three targets are depicted in Figure \ref{O-C}. Since the time span of the eclipsing times isn't long enough, more observations will be needed in the future, especially for J034828.

\begin{figure}
	\epsscale{0.36}
	\plotone{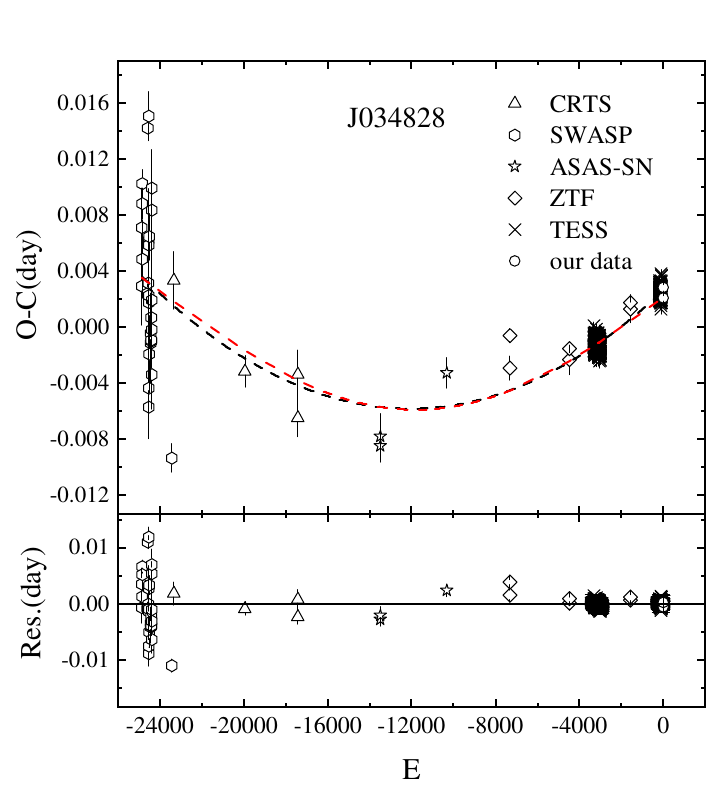}
	\plotone{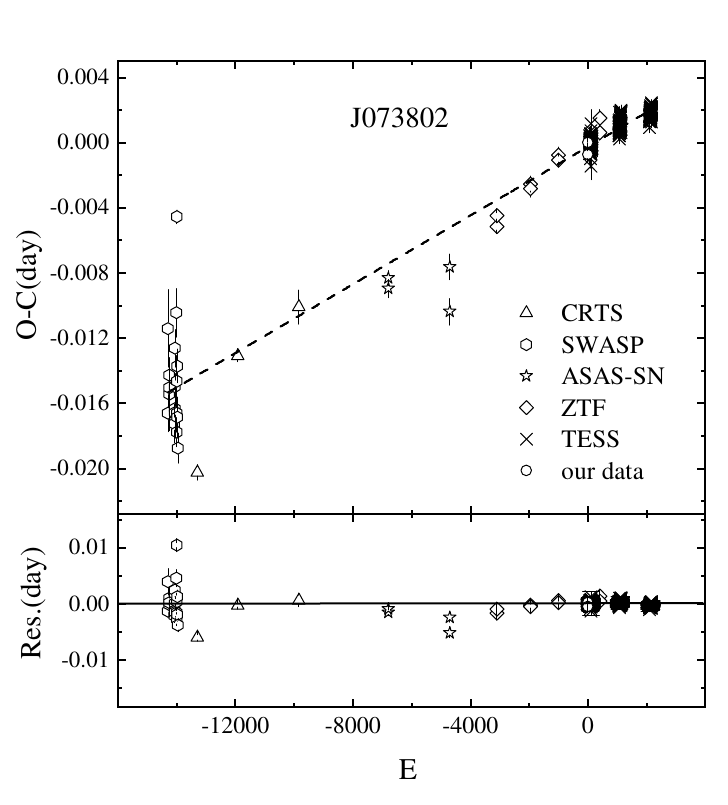}
	\plotone{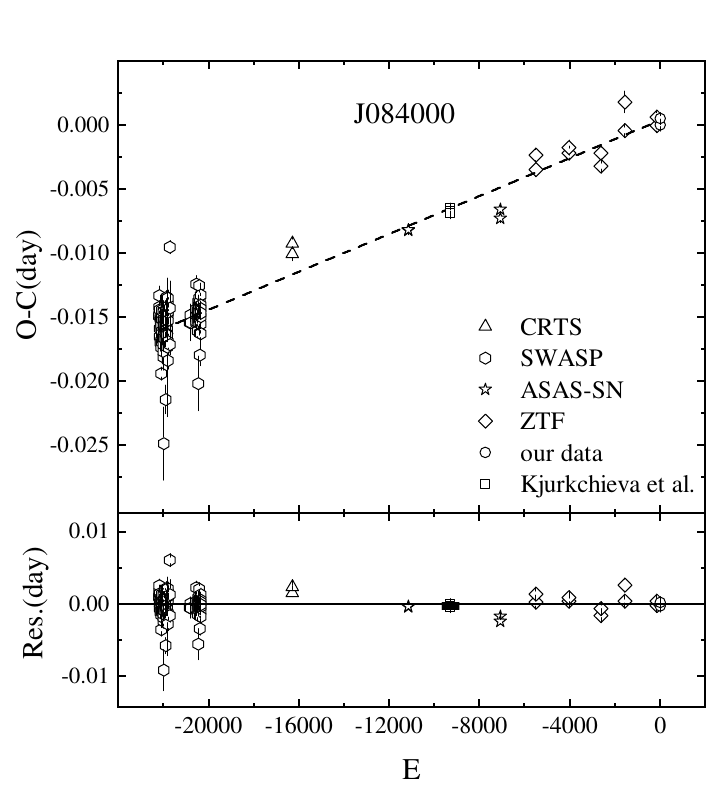}
	
	\caption{The O-C diagrams of the three targets. For diagram of J034828, the black line represents a parabolic fit, and the red line represents a periodic sine function fit.
	\label{O-C}}
\end{figure}
%表8

\begin{deluxetable*}{ccccccc}
	\tablecaption{The eclipsing times and O-C values of the three targets.
		\label{tab:O-C1}}
	\tablenum{9}
	\tablehead{\colhead{Target} & \colhead{BJD} & \colhead{Error} & \colhead{E} & \colhead{O-C} & \colhead{Residual} & \colhead{Telescope} } 
	\startdata
	J034828 & 2454002.65492 & 0.00277  & -24852 & 0.00292  & -0.00069  & SWASP \\
	& 2454003.67026 & 0.00283  & -24848 & 0.00710  & 0.00349  & SWASP \\
	& 2454004.67920 & 0.00279  & -24844 & 0.00487  & 0.00127  & SWASP \\
	& 2454005.69574 & 0.00100  & -24840 & 0.01024  & 0.00664  & SWASP \\
	& 2454006.70549 & 0.00112  & -24836 & 0.00882  & 0.00523  & SWASP \\
	& 2454070.40271 & 0.00068  & -24584 & 0.00238  & -0.00084  & SWASP \\
	& 2454071.42573 & 0.00044  & -24580 & 0.01423  & 0.01100  & SWASP \\
	& 2454077.60804 & 0.00096  & -24555.5 & 0.00313  & -0.00005  & SWASP \\
	& 2454083.42087 & 0.00282  & -24532.5 & 0.00173  & -0.00141  & SWASP \\
	& 2454083.55174 & 0.00345  & -24532 & 0.00621  & 0.00306  & SWASP \\
	& 2454084.42458 & 0.00120  & -24528.5 & -0.00572  & -0.00886  & SWASP \\
	\enddata
	\tablecomments{ This table is available in its entirety in machine-readable form in the online version of this article. }
\end{deluxetable*}

%表9
\begin{deluxetable}{ccccc}
	\tablecaption{The corrected values of initial epoch and period of the three targets.
		\label{tab:O-C2}}
	\tablenum{10}
	\tablehead{\colhead{Target} & \colhead{$\Delta T_0$} & \colhead{Error} & \colhead{$\Delta P$} & \colhead{Error} \\ 
		\colhead{} & \colhead{($10^{-4}$d)} & \colhead{($10^{-4}$d)} & \colhead{($10^{-6}$d)} & \colhead{($10^{-6}$d)} } 
	\startdata
	J034828  &  -20.60  &  2.46  &  -0.30  &  0.03  \\
	J073802  &  -2.54  &  0.48  &  1.05  &  0.13  \\
	J084000  &  3.44  &  4.16  &  0.73  &  0.02  \\
	\enddata
\end{deluxetable}
\section{Spectroscopic Investigation}
From the asymmetric light curves we observed, we can infer that our three targets have strong  photospheric magnetic activity. For chromospheric activity in late type stars, the emission lines such as $H_\alpha$ line can serve as reliable indicators \citep{Halpha}. However, the observed spectrum of a star typically shows absorption in the H$\alpha$ line, which is attributed to the absorption in the stellar photosphere. Thus we employed the spectral subtraction method \citep{spectra} to identify the emission line. Initially, we selected two inactive stars spectra (according to the standard that the temperature difference between the inactive star and the target is less than 200K) from the catalog of \citet{spectralsubtraction}, one of the two spectra served as the primary template, while the other served as the secondary template, after that we downloaded and normalized the spectra. Utilizing the STARMOD code \citep{spectra}, we generated synthetic spectra for the two inactive stars accounting for radial velocities, spin angular velocities, and the luminosity ratios of the two components. Then we obtained the subtracted spectra from the synthetic spectra and the target spectra, all of them are plotted in Figure \ref{spectra}. Notably, all the three targets exhibit $H_\alpha$ emission lines, indicating they all have chromospheric activity. Finally we calculated the equivalent widths (EWs) of the $H_\alpha$ line using the splot package of the IRAF\footnote{ \url{https://iraf.noirlab.edu/}} to further quantify the chromospheric activity.  It is found that all the three targets have EWs more than 0.6\AA, indicating strong chromospheric activity of them.

\begin{figure}
	\epsscale{0.36}
	\plotone{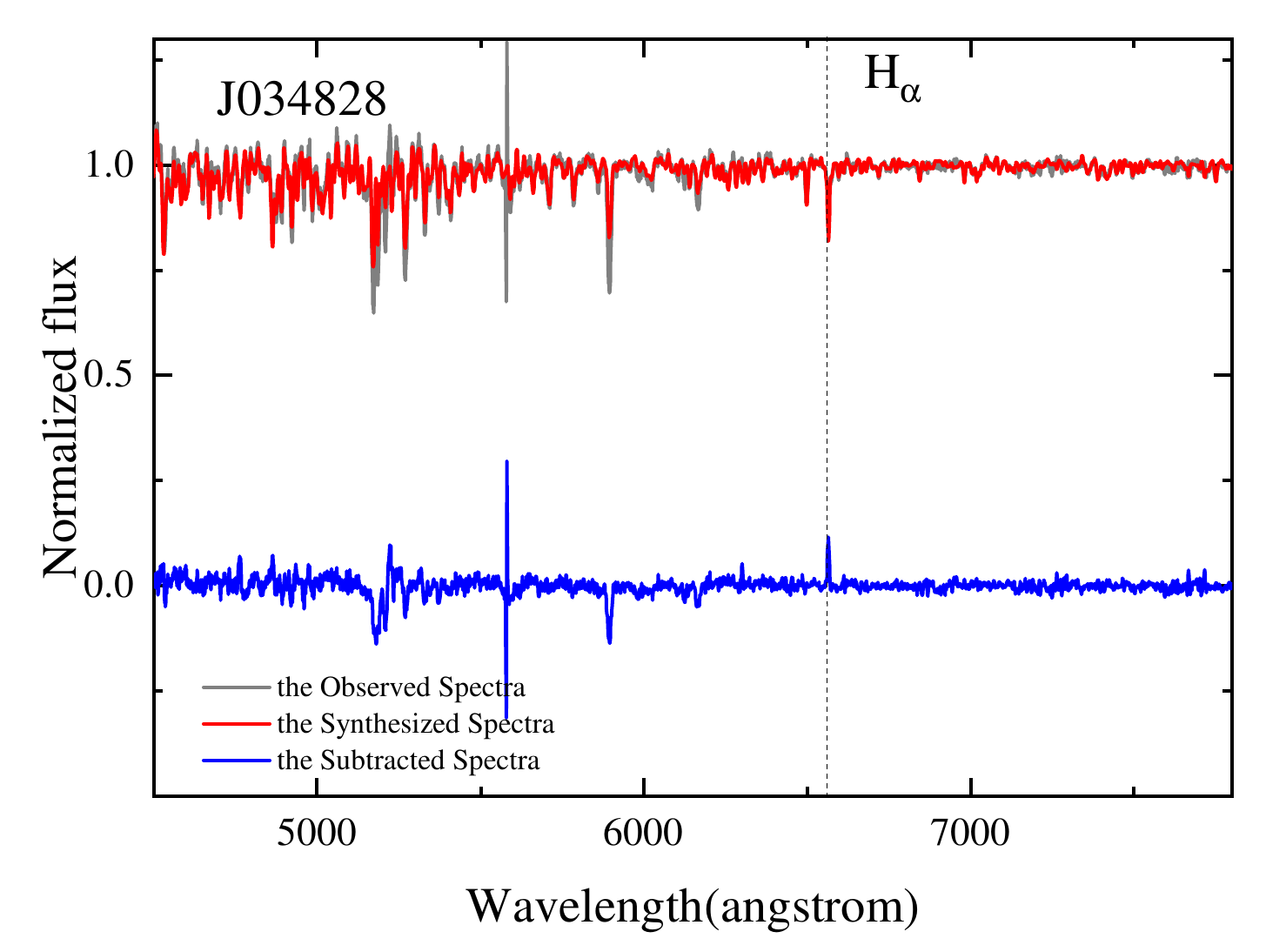}
	\plotone{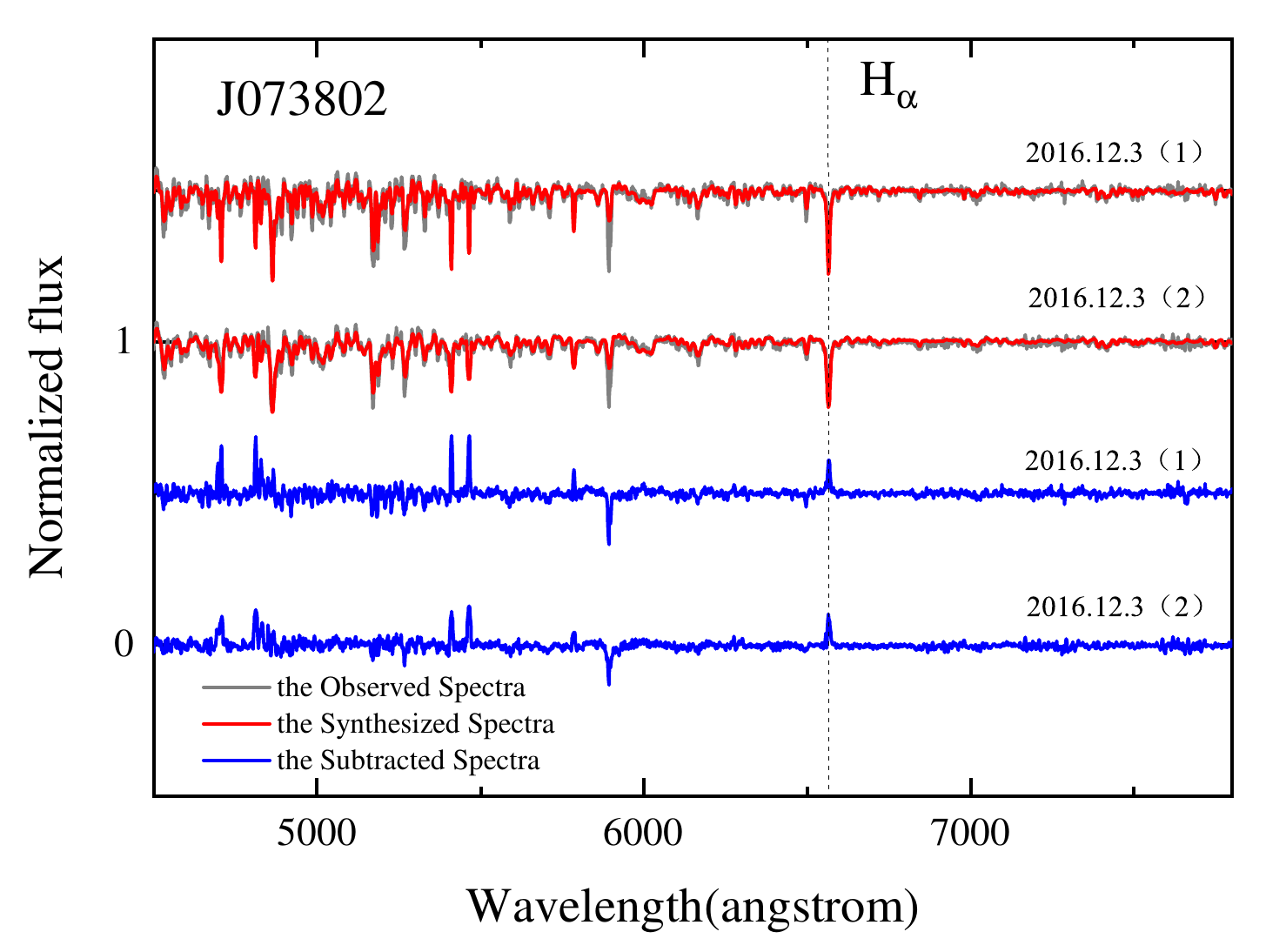}
	\plotone{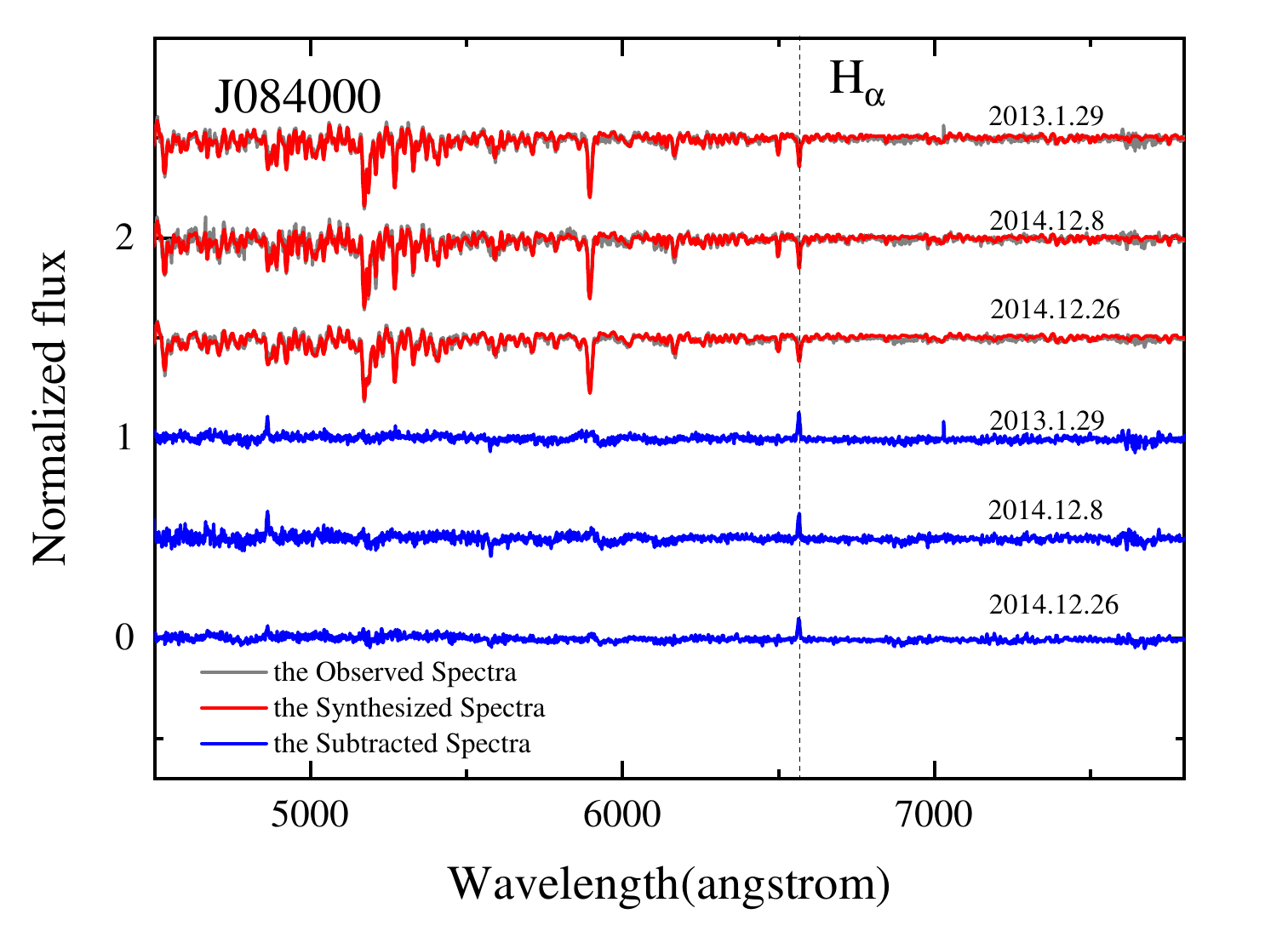}
	\caption{The spectroscopic investigations for the three targets, including one for J034828, two for J073802, and three for J084000.
	\label{spectra}}
\end{figure}

\section{Discussion and Conclusion}
We presented  photometric and spectroscopic investigation of three large amplitude contact binaries. The previous studies found nearly equal mass ratios for totally eclipsing contact binaries in photometric and spectroscopic methods. \citep{lightcurve, Terrell, Gaia}. \citet{2024arXiv240600155R} raised doubt on the Lucy model and the photometric mass ratio. However, due to the absence of radial velocity curves, we determined the mass ratio merely on the basis of photometric observations. Given that the orbital inclinations of the three targets are all greater than 85\degr, and we believe they are not affected by a third light, the photometric mass ratios remain relatively reliable.
 During the light curves analysis, we found that they all have O'Connell effect and a cool spot was added on the primary component to get a better fit. It's corresponding to our spectroscopic analysis results, revealing intense magnetic activity of the three targets. It was also found that J084000 can be classified as A-subtype contact binaries, J034828 and J073802 can be classified as W-subtype contact binaries. Furthermore, our photometric solution indicated all the three targets belong to shallow contact binaries and confirmed that J084000 belongs to H-subtype contact binary.
\par We compared our photometric solutions with the results of previous work. The physical parameters obtained by previous work are listed in Table \ref{previous work}. For J073802, the difference between the parameters determined by \citet{11000} and ours is enormous, especially the mass ratio. We analyzed its  ASAS-SN light curves and the result is corresponding to ours but far from \cite{11000}. Since \citet{11000} employed the machine learning method to obtain physical parameters, which may lead to deviations in the results, the solutions we determined are more reliable. Our results for J084000 are in agreement with those of \citet{11000}; however, a clear distinction arises when these results are compared to those of \citet{2017RMxAA..53..133K}. Comparing our light curves with the light curves they observed, it is found that theirs exhibit stronger O'Connell effect. Then we utilized the q-search method to analyze Kjurkchieva's light curves and  plotted the relationship between the mean residuals and mass ratios in Figure \ref{qsearch2}. The minimum mean residual was found at q=0.8 which is corresponding to our results, indicating the difference in physical parameters can primarily be attributed to the impact of star spots \citep{Liuliang2021PASP..133h4202L}. Given that the light curves we observed are less influenced by star spots, we considered our results are more reliable.

\vspace{-0.8cm}

\begin{deluxetable*}{cccccc}
	\tablecaption{The physical parameters we obtained and from previous work.
		\label{previous work}}
	\tablenum{11}
	\tablehead{\colhead{Target} & \multicolumn2c{J073802}  & \colhead{} & \colhead{J084000} & \colhead{}\\ 
		\colhead{Reference} & \colhead{Ours} & \colhead{Li et al.} & \colhead{Ours} & \colhead{Kjurkchieva et al.} & \colhead{Li et al.}} 
	
	\startdata
	$T_1$(K) & 5651$\pm$56 & 5528 & 4910$\pm$71 & 4689$\pm$25 & 4927 \\
	$i(\degr)$ & 89.7$\pm$0.1 & 84.2$\pm$0.7 & 85.3$\pm$0.1 & 84.9$\pm$0.4 & 84.3$\pm$1.4 \\
	$q$ & 0.668$\pm$0.003 & 0.99$\pm$0.23 & 0.78$\pm$0.005 & 0.9$\pm$0.1 & 0.73$\pm$0.17 \\
	$f$ & 10.0\%$\pm$1.3\% & 3.3\%$\pm$1.2\% & 14.4\%$\pm$2.6\% & 10.90\% & 10\%$\pm$3\% \\
	$T_2$(K) & 5712$\pm$94 & 5499 & 4798$\pm$113 & 4613$\pm$24 & 4927 \\
	Subtype & W & A & A & A & A \\
	$r_1$ & 0.424$\pm$0.001 & 0.384 & 0.414$\pm$0.001 & 0.399$\pm$0.007 & 0.417 \\
	$r_2$ & 0.352$\pm$0.099 & 0.382 & 0.371$\pm$0.003 & 0.381$\pm$0.007 & 0.360 \\
	$L_2/L_1$ & 0.728$\pm$0.002 & 0.968 & 0.719$\pm$0.005 & 0.837 & 0.745 \\
	\enddata
	
\end{deluxetable*}

\begin{figure}
	\centering
	\includegraphics[width=0.5\textwidth]{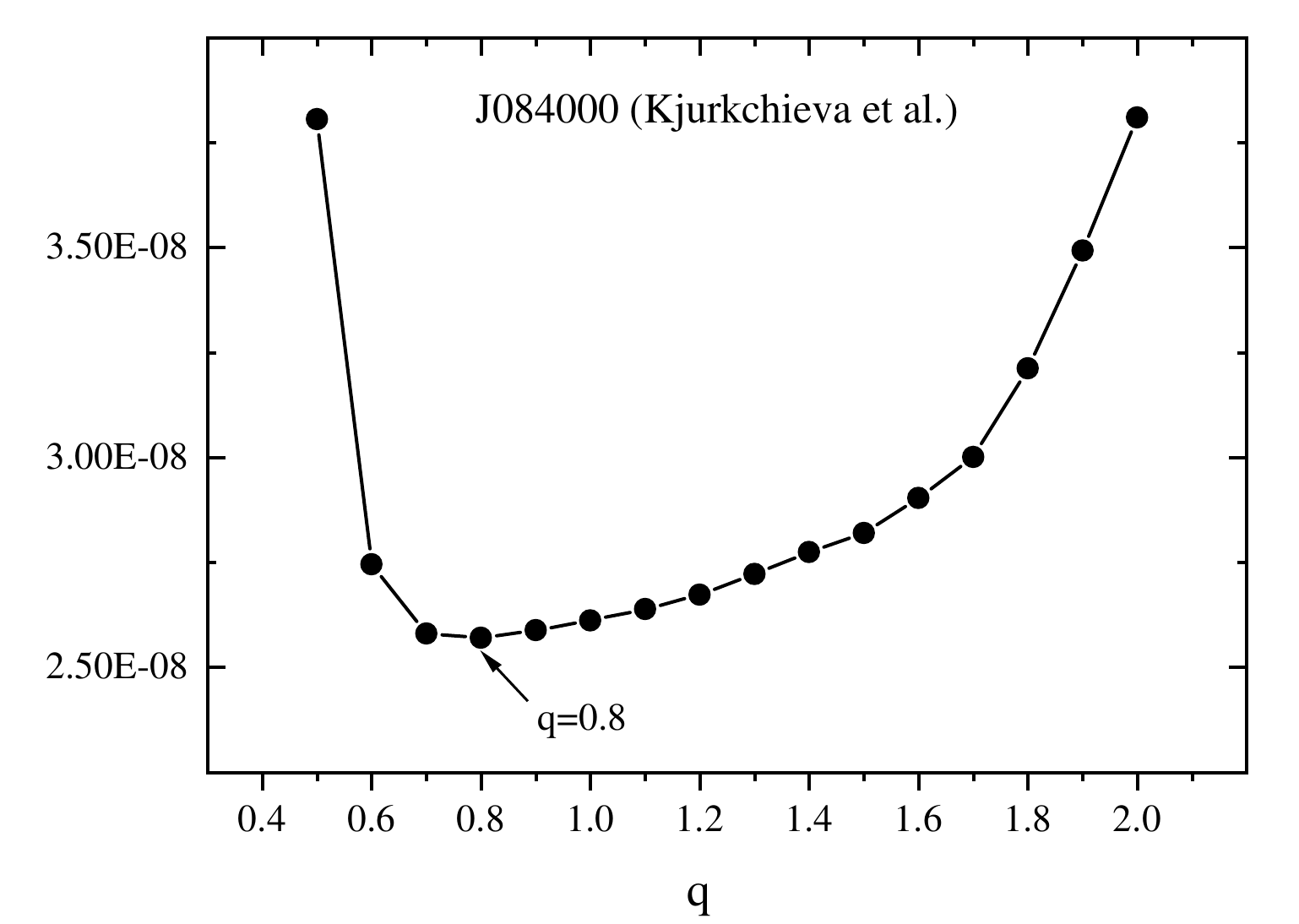}
	\caption{The relationship between mean residuals and mass ratio of light curves from \citet{2017RMxAA..53..133K}.}
	\label{qsearch2}    
\end{figure}
\par Based on the investigation conducted by \citet{Gaia} , we can utilize Gaia distance to estimate the absolute parameters. We used $M_v = m_{vmax}-5logD+5-A_v$  to derive the absolute magnitudes ($Mv$), where $A_v$ is the extinction coefficients, $m_{vmax}$ is the brightest V-band magnitude obtained from the ASASSN theoretical curve, D is the distance from {\it Gaia} DR3. Then we obtained the bolometric corrections ($BC_v$) using the interpolation from the table provided by \citet{BCv}. The absolute bolometric magnitude $M_{bol}$ can be calculated using the equation $M_{bol}=M_{v}+BC_v$. According to $M_{bol}=-2.5lgL_{total}+4.75$, we estimated the luminosities of the primary and secondary components ($L_{1,2}$). The equation $M_{\mathrm{bol}}=-2.5\mathrm{lg}((\frac{A\times r_{1}}{R_{\odot}})^{2}(\frac{T_{1}}{T_{\odot}})^{4} + (\frac{A\times r_{2}}{R_{\odot}})^{2}(\frac{T_{2}}{T_{\odot}})^{4})) + 4.75$ was used to determine the distance between the two components (A),
where $r_1$ and $r_2$ is the relative radii of primary and secondary components obtained from the photometric solutions. Based on $R_{1}/R_{\odot}=r_{1}\times A$ and $R_{2}/R_{\odot}=r_{2}\times A$ \citep{R1R2}, the primary and secondary radii ($R_{1,2}$) in solar units can be obtained. Ultimately, utilizing the Kepler's third law $M_1 + M_2 = 0.0134\frac{A^3}{P^2}$, where P is in days, A is in solar units ($R_\sun$), we obtained the total mass of the system. Then using the mass ratio obtained from the W-D program, we calculated the masses of primary and secondary ($M_{1,2}$). The absolute parameters and errors are listed in Table \ref{absolute parameters}.
\par \citet{q=0.72} defined H-subtype (q\textgreater0.72) contact binaries which have less efficient energy transfer rate at a given luminosity ratio. In this paper we collected 52 H-subtype contact binaries, the basic parameters of them are shown in Table \ref{H-subtype}. Then we studied the energy transfer behaviors of 76 H-subtype systems, including 52 objects collected by us and 24 objects analyzed by \citet{q=0.72} previously. The energy transfer parameter can be defined as  \citep{q=0.72}:
\begin{equation}
	\beta=\frac{L_{1,observed}}{L_{1,ZAMS}},
\end{equation}
where $L_{1,observed}$ is the observed luminosity of primary and $L_{1,ZAMS}$ is the luminosity of the primary corresponding to ZAMS. So according to \citet{q=0.72}, it is easy to calculate that 
\begin{eqnarray}
		\beta=\frac{1+q^{4.216}}{1+q^{0.92}\left(\frac{T_{2}}{T_{1}}\right)^{4}}=\frac{1+\alpha\lambda^{4.58}}{1+\lambda},  
\end{eqnarray}
where $\alpha=(\frac{T_1}{T_2})^{18.3}$, $\lambda$ is the bolometric luminosity ratio. Here the indices are corrected by \citet{Sun2020ApJS} according to M-L relation provided by \citet{2013MNRAS.430.2029Y}. 
\par The relationships between energy transfer parameter $\beta$ and luminosity ratio $\lambda$ of 76 H-subtype contact binaries are shown in Figure \ref{beta}. It is found that A-subtype systems exhibit higher transfer parameters compared to W-subtype systems. Furthermore, all systems are encompassed by an envelope that represents the minimum transfer parameter at a given luminosity ratio ($\alpha$ = 0).The following equation can be used to compute this minimum rate \citep{q=0.72},
\begin{equation}
	\beta_{envelope}=\frac{1}{1+\lambda}.
\end{equation}
Compared to the non-H contact binaries cataloged by \citet{Gaia}, H-subtype contact binaries are located far from this envelope. This unique energy transfer behavior can be explained by the near-equal masses of the two components, which results in a smaller luminosity transfer. What's more, according to the distribution of H-subtype in the diagram with $\alpha$ ranging from 0.5 to 8, it can be deduced that for high mass ratio contact binaries, the temperatures of two components can be very different. In fact, \citet{q=0.72} corrected the $\beta$ values in their paper, where $\beta_{corr}=\beta-0.52q^{4.1}$. We plotted the relationship between the corrected transfer parameter $\beta_{corr}$ and luminosity ratio $\lambda$ in Figure \ref{beta}. Almost all the systems have a good correlation between the corrected transfer parameter and luminosity ratio, so we confirmed the conclusion of them. We have plotted the dependence of the energy transfer parameter $\beta$ and $\beta{corr}$ on the contact degree ($f$) in Figure \ref{betanew}. A positive correlation has been observed in the two panels, indicating that the energy transfer parameter increases with the contact degree. We used the following equation to fit them,
\begin{eqnarray}
	\beta&=0.11(\pm0.03)f+0.75(\pm0.01) 
	\nonumber \\
	\beta_{corr}&=0.20(\pm0.01)f+0.63(\pm0.01).
\end{eqnarray}
	 In the $\beta_{corr}-f$ diagram, it's evident that for objects with the same contact degree, the H subtype systems display a lower corrected transfer parameter compared to the non-H objects.

\begin{figure}
	\epsscale{0.5}
	\plotone{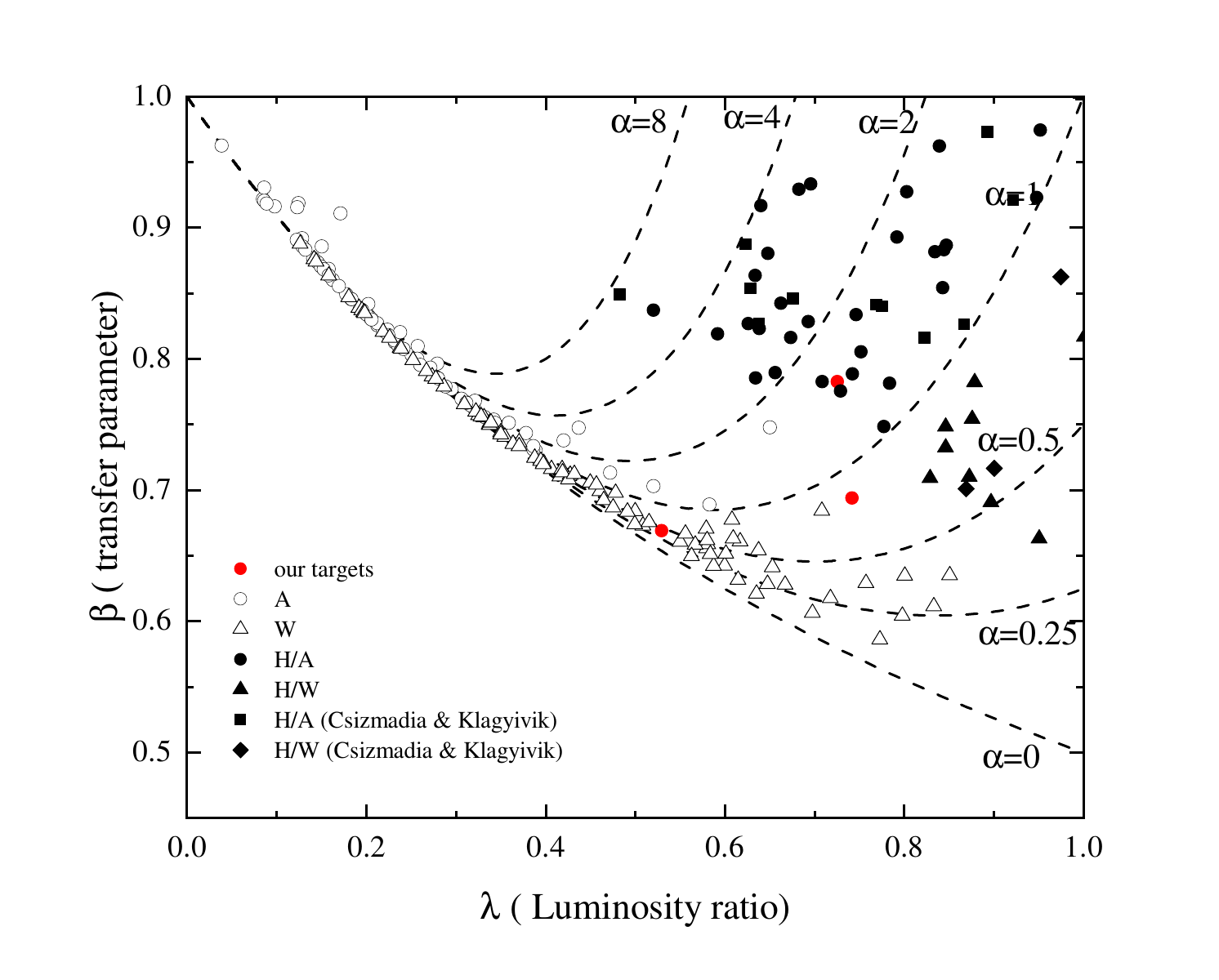}
	\plotone{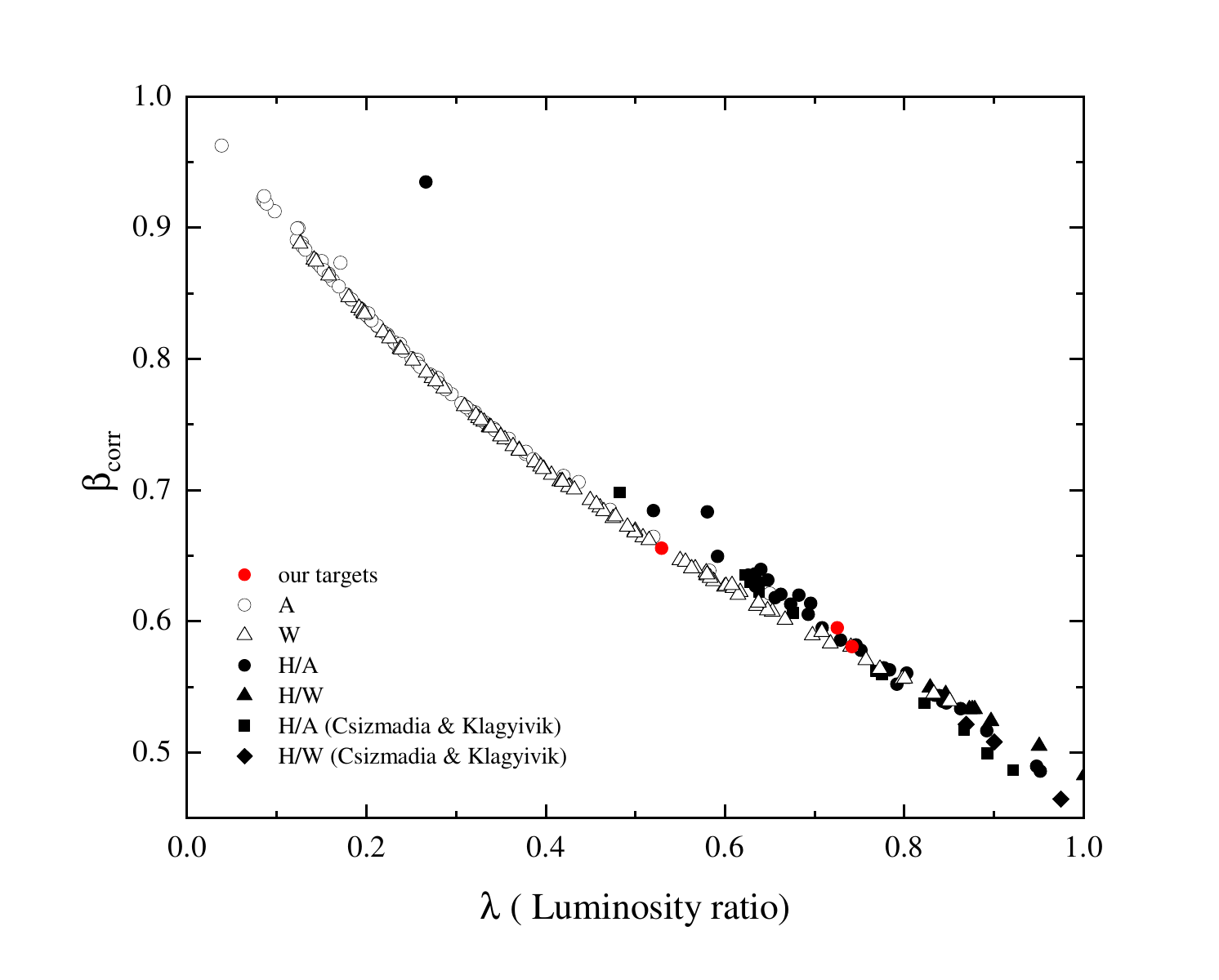}
	\caption{The left is energy transfer parameter–bolometric luminosity ratio diagram and the right is the corrected transfer parameter-bolometric luminosity diagram. The dashed lines correspond to different $\alpha$ values. The non-H systems are collected by \citet{Gaia}, which are depicted as open circles and triangles for A and W subtype. The filled circles and triangles depict the H/A and H/W contact binaries we collected. 
		The filled squares and diamonds depict the H/A and H/W systems collected by \citet{q=0.72}. 
	\label{beta} }   
\end{figure}

\begin{figure}
	\epsscale{0.5}
	\plotone{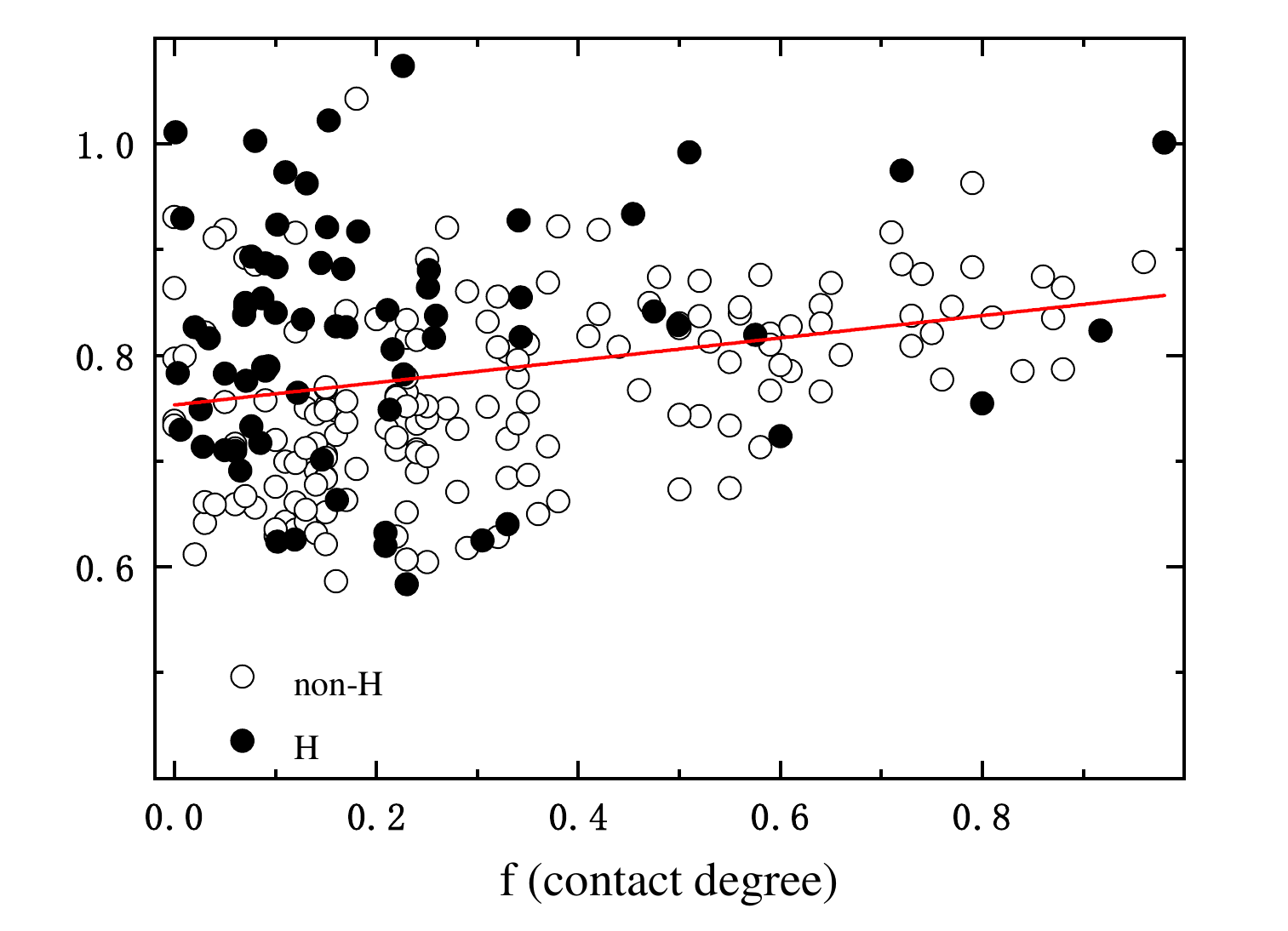}
	\plotone{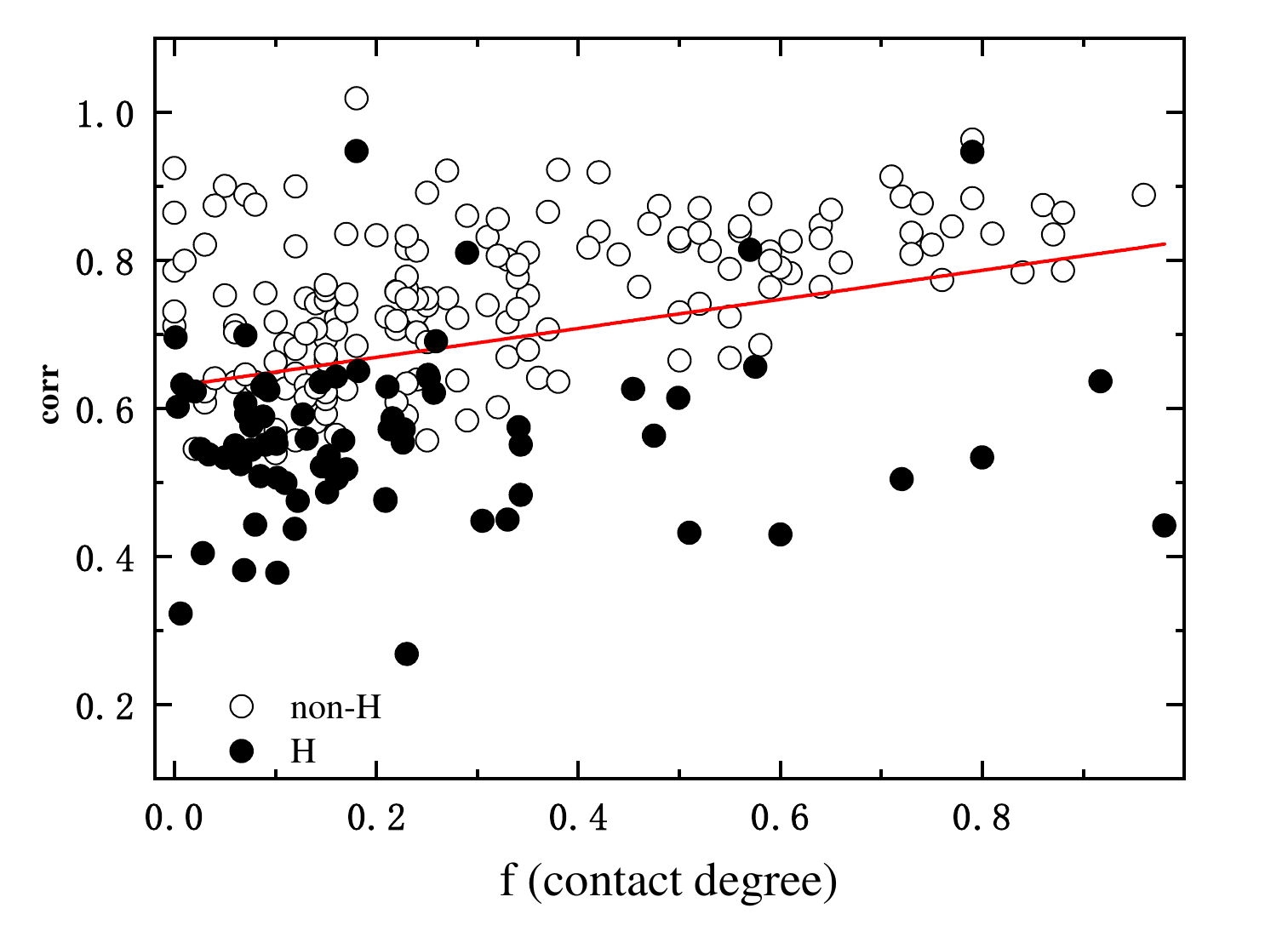}
	\caption{The left is energy transfer parameter–contact degree diagram and the right is the corrected transfer parameter-contact degree diagram.
		\label{betanew} }   
\end{figure}

\par According to TRO, a contact binary will oscillate between  contact-broken phase and contact-phase. At the contact stage, if the mass is transferring from the less massive component to the more massive one, it will lead a decreasing mass ratio. For high mass ratio, shallow contact degree contact binaries such as J073802 and J084000, they may be at the beginning of the contact phase and are dominated by TRO. For J034828 which has a lower mass ratio, the results of its orbital period analysis showed that it exhibits a long-term period increase which may be caused by mass transfer. The following equation can be used to calculate the mass transfer rate \citep{1958BAN....14..131K},
\begin{equation}
	\frac{\dot{P}}{P}=-3\dot{M}_1\biggl(\frac{1}{M_1}-\frac{1}{M_2}\biggr).
\end{equation}
We determined that the less massive component is transferring mass to the more massive one at a rate of $1.38\pm 0.04\times10^{-7}M{_\sun} yr^{-1}$. As the mass ratio decreases, a critical transition occurs when q falls below 0.79. During this phase, both the primary and secondary Roche-lobe radii expand, eventually exceeding the radii of two components. This expansion ultimately leads to  a contact binary system evolve into a semi-detached binary, and the mass transfer will stop \citep{2019AJ}. Therefore, the shallow contact binary J034828 with an increasing period is  controlled by TRO and it is evolving into a broken-contact phase.

\begin{deluxetable*}{ccccccccccc}
	\tablecaption{The estimated absolute parameters and the orbital angular momentum of the three targets.
		\label{absolute parameters}}
	\tablenum{12}
	\tablehead{\colhead{Target} & \colhead{$T_1$(K)} & \colhead{$T_2$(K)} & \colhead{a($R_\sun$)} & \colhead{$M_1$($M_\sun$)} & \colhead{$M_2$($M_\sun$)} & \colhead{$R_1$($R_\sun$)} & \colhead{$R_2$($R_\sun$)} & \colhead{$L_1$($L_\sun$)} & \colhead{$L_2$($L_\sun$)} & \colhead{$J_o$(cgs)}  }
	\startdata
	J034828 & 5059±58 & 4831±81 & 1.80±0.03 & 0.36±0.02 & 0.87±0.05 & 0.57±0.01 & 0.85±0.02 & 0.19±0.01 & 0.35±0.01 & 2.271$\times$ $10^{51}$ \\
	J073802 & 5651±56 & 5712±94 & 2.64±0.05 & 1.15±0.07 & 0.77±0.05 & 1.12±0.02 & 0.93±0.02 & 1.14±0.04 & 0.84±0.03 & 6.232$\times$ $10^{51}$ \\
	J084000 & 4910±71 & 4798±113 & 1.95±0.03 & 0.79±0.04 & 0.62±0.03 & 0.81±0.01 & 0.72±0.02 & 0.34±0.01 & 0.24±0.01  & 3.493$\times$ $10^{51}$ \\
	\enddata
\end{deluxetable*}

\par To gain a deeper understanding of the evolutionary status of the three contact binaries, we plotted mass-luminosity (M–L) and mass-radius (M–R) diagrams of three targets and collected contact binaries in Figure \ref{ML}. In this figure, the zero-age main sequence (ZAMS) and terminal-age main sequence (TAMS) lines are provided by \citet{ZAMS} and depicted by solid and dashed lines. Here the primary component in each system is the more massive component. Our analysis reveals that the primary components are distributed around the ZAMS line, indicating that they are slightly evolved stars. Apparently, the secondary are more evolved than the primary. It can be attributed to the energy being transferred from the primary to the secondary component. It's corresponding to the previous studies \citep{evolution,Gaia}.
For all systems with a higher mass ratio, both the primary and secondary components lie on the main sequence between ZAMS and TAMS lines. However, for J034828 which has a mass ratio lower than 0.5, the secondary component is positioned above the TAMS line. This indicates that the secondary component of J034828 is over-luminous and has evolved out of the main sequence.

\begin{figure}[t]
	\epsscale{0.5}
	\plotone{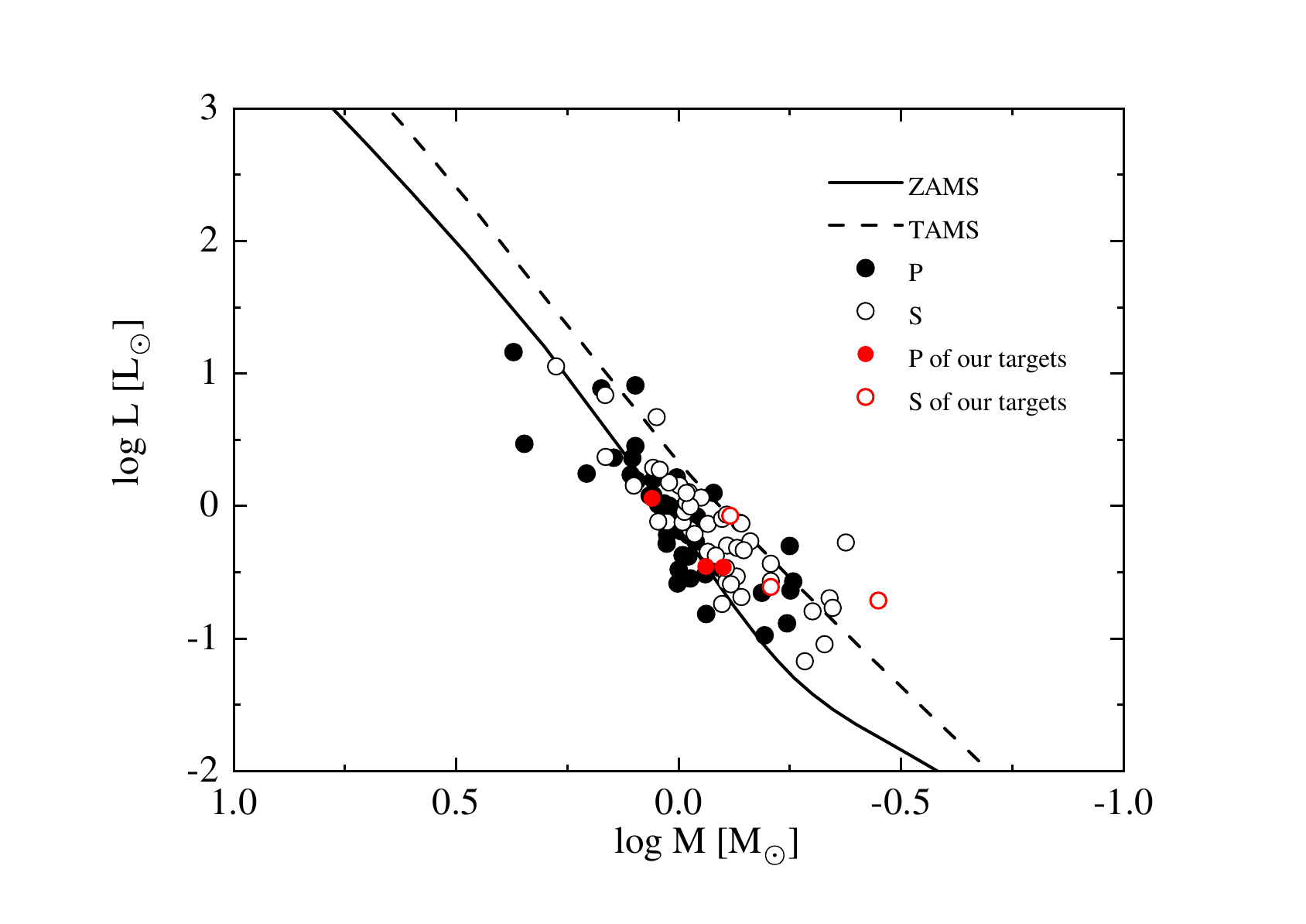}
	\plotone{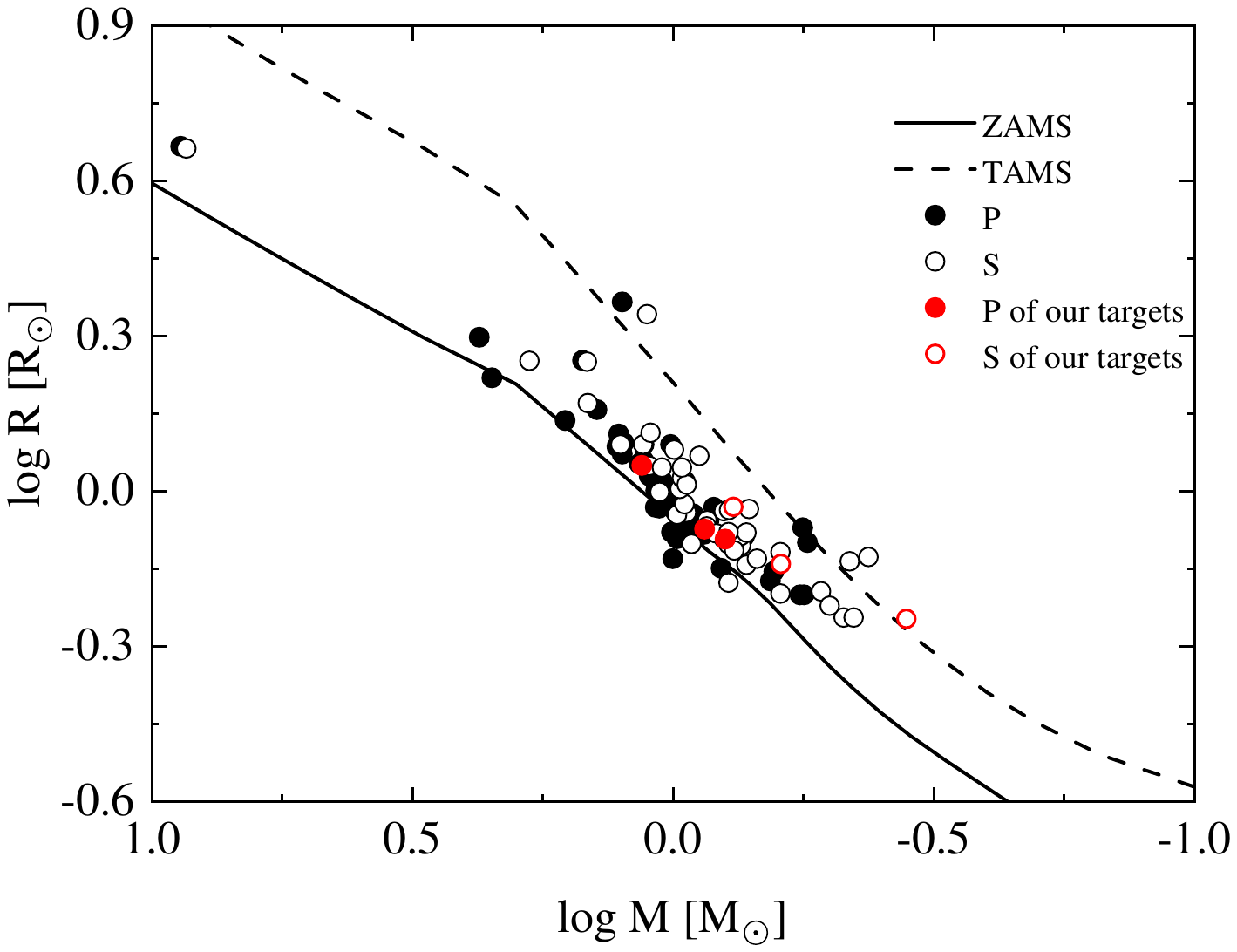}
	\caption{The mass–luminosity and mass–radius diagrams, the left is M-L diagram and the right is M-R diagram. The black solid and open circles represent primary and secondary of collected H-subtype targets and the red solid and open circles represent primary and secondary of our targets. 
	\label{ML} }   
\end{figure}
\begin{figure}
	\centering
	\includegraphics[width=0.5\textwidth]{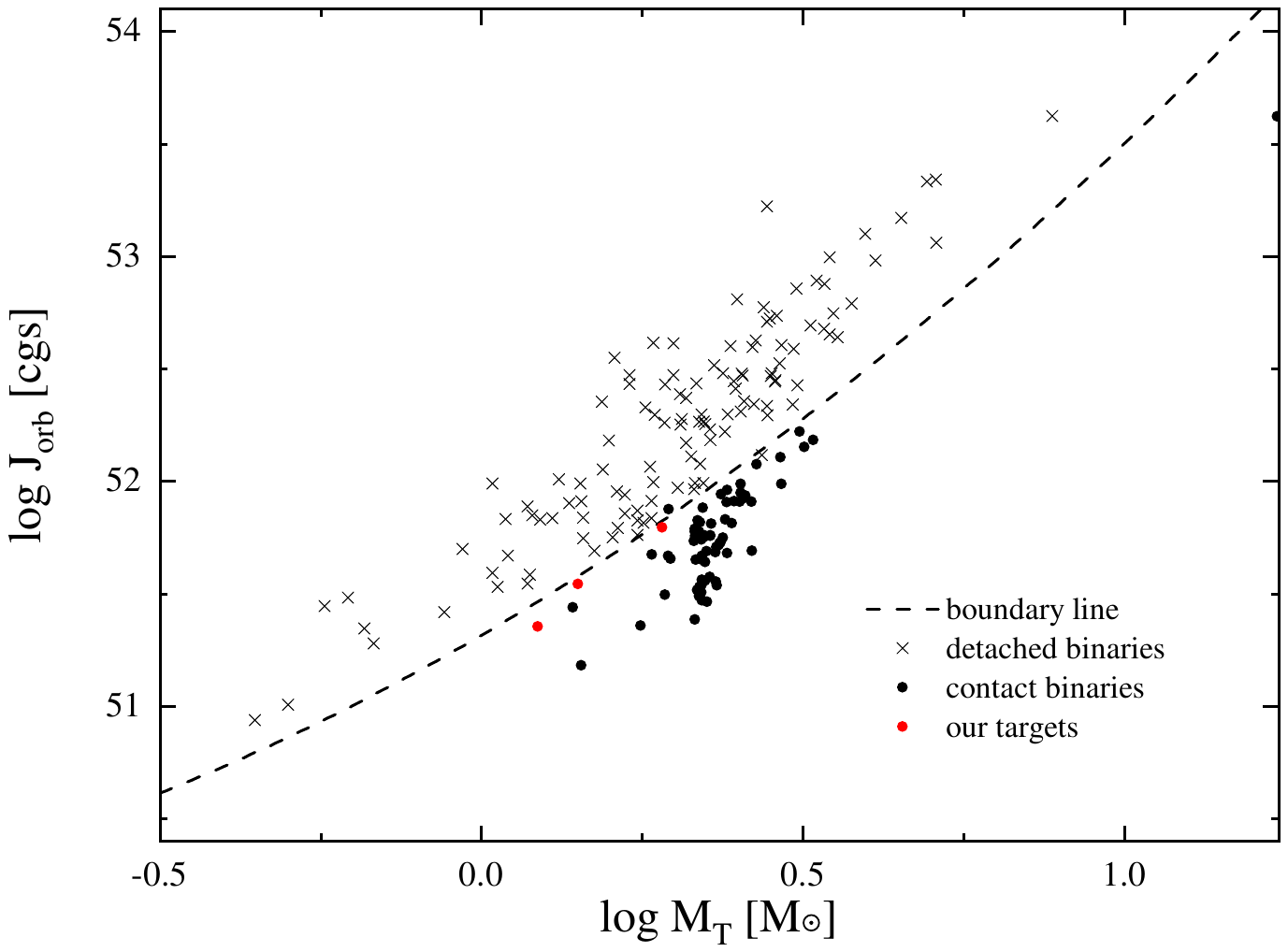}
	\caption{The log$J_o$-log$M_{total}$  diagram. The black crosses represent the detached binaries collected by \citet{Jodiagram}. The black solid  circles represent the contact binaries collected by \citet{Gaia}. The solid red circles are our three targets. This diagram also incorporates a borderline depicted as a dashed line.
		\label{Jo}}
\end{figure}

 The orbital angular momentum ($J_o$) can be calculated utilizing the following equation \citep{Jo}:
\begin{equation}
	J_o=1.24\times10^{52}\times M_{\mathrm{total}}^{\frac53}\times P^{\frac13}\times\frac q{(1+q)^2},
\end{equation}
where $M_{total}=M_1+M_2$, it's the total mass of the contact binaries systems, P is the orbital period and q is the mass ratio. The results of our targets are listed in Table \ref{absolute parameters}. The log$J_o$-log$M_{total}$  diagram serves as a valuable tool in investigating the evolution of contact binary stars. We plotted the diagram of our three targets in Figure \ref{Jo}. The dashed line is the contact border identified by \citet{Jodiagram}, which effectively distinguishes detached systems from contact binaries. All the targets are located below the boundary line, with J073802 and J084000 relatively close to it. These positions indicates that they possess greater angular momentum compared to contact binaries of the same total mass. This finding support the conclusion that they are in the early stages of the contact phase and  follow the TRO. We can also infer that these contact binaries have evolved from short-period detached binaries via angular momentum loss (AML).
\par In conclusion, we analyzed three large amplitude contact binaries, obtained the physical parameters of them and studied their evolutionary states. All of them are determined as shallow contact binaries. We presented the O-C diagrams of the three targets using available eclipsing times and studied the mass transfer of J034828. We determined the absolute parameters using the photometric solutions and Gaia distance.The spectral subtraction method was used to investigate their chromospheric activities, they all exhibit strong $H_\alpha$ emission lines. Future observations, especially the radial velocity observations, are needed to determine precise absolute parameters.

\begin{acknowledgments}
	
	\par Thanks the anonymous referee very much for the constructive and insightful criticisms and suggestions to improve our manuscript. This work is supported by National Natural Science Foundation of China (NSFC) (No.12273018), and the Joint Research Fund in Astronomy (No.U1931103) under cooperative agreement between NSFC and Chinese Academy of Sciences (CAS), and by the Qilu Young Researcher Project of Shandong University, and by the Young Data Scientist Program of the China National Astronomical Data Center, and by the Cultivation Project for LAMOST Scientific Payoff and Research Achievement of CAMS-CAS, and by the Instrument Education Funds of Shandong University (yr20240205). The calculations in this work were carried out at Supercomputing Center of Shandong University, Weihai.
	\par The photometric observations were conducted using  Weihai Observatory 1.0-m telescope of Shandong University (WHOT), the 60 cm telescope at the Xinglong Station of National Astronomical Observatories (XL60) and the 60 cm Ningbo Bureau of Education and Xinjiang Observatory Telescope (NEXT), we acknowledge the support of the staff of them. 
	\par The spectra observations were provided by Guoshoujing Telescope (the Large Sky Area Multi-Object Fiber Spectroscopic Telescope; LAMOST), which is a National Major Scientific Project built by the Chinese Academy of Sciences. Funding for the project has been provided by the National Development and Reform Commission. LAMOST is operated and managed by the National Astronomical Observatories, Chinese Academy of Sciences. 
	\par We thank Las Cumbres Observatory and its staff for their continued support of ASAS-SN. ASAS-SN is funded in part by the Gordon and Betty Moore Foundation, United States through grants GBMF5490 and GBMF10501 to the Ohio State University, and also funded in part by the Alfred P. Sloan Foundation grant G-2021-14192. 
	\par Based on observations obtained with the Samuel Oschin 48 inch Telescope at the Palomar Observatory as part of the Zwicky Transient Facility project. ZTF is supported by the National Science Foundation under grant No. AST-1440341 and a collaboration including Caltech, IPAC, the Weizmann Institute for Science, the Oskar Klein Center at Stockholm University, the University of Maryland, the University of Washington, Deutsches Elektronen-Synchrotron and Humboldt University, Los Alamos National Laboratories, the TANGO Consortium of Taiwan, the University of Wisconsin at Milwaukee, and Lawrence Berkeley National Laboratories. Operations are conducted by COO, IPAC, and UW.
	\par This paper makes use of data from the DR1 of the WASP data \citep{SWASP} as provided by the WASP consortium, and the computing and storage facilities at the CERIT Scientific Cloud, reg. no. CZ.1.05/3.2.00/08.0144, which is operated by Masaryk University, Czech Republic.
	\par This research was made possible through the use of the AAVSO Photometric All-Sky Survey (APASS), funded by the Robert Martin Ayers Sciences Fund and NSF AST-1412587.
	\par This work makes use of data collected by the TESS mission. Funding for the TESS mission is provided by NASA Science Mission directorate. We acknowledge the TESS team for its support of this work.
	\par This paper makes use of data from the Two Micron All Sky Survey (MASS), which is a joint project of the University of Massachusetts and the Infrared Processing and Analysis Center/California Institute of Technology. Funding of MASS is provided by the National Aeronautics and Space Administration and the National Science Foundation.
	\par This work has made use of data from the European Space Agency (ESA) mission
	{\it Gaia} (\url{https://www.cosmos.esa.int/gaia}), processed by the {\it Gaia} Data Processing and Analysis Consortium (DPAC,
	\url{https://www.cosmos.esa.int/web/gaia/dpac/consortium}). Funding for the DPAC
	has been provided by national institutions, in particular the institutions
	participating in the {\it Gaia} Multilateral Agreement.
	\par The CRTS survey is supported by the US National Science Foundation under grants AST-0909182, AST-1313422, AST1413600, and AST-1518308.
	
\end{acknowledgments}

\appendix

\begin{longrotatetable}
	\begin{deluxetable*}{cccccccccccccc}

		\tablecaption{The physical parameters of 52 H-subtype contact binaries.
			\label{H-subtype}}
		\tablewidth{0pt}
		\tablenum{A1}
		\tablehead{\colhead{Name} & \colhead{Type} & \colhead{P} & \colhead{q} & \colhead{f} & \colhead{$T_1$} & \colhead{$T_2$} & \colhead{$M_1$} & \colhead{$M_2$} & \colhead{$R_1$} & \colhead{$R_2$} & \colhead{$L_1$} & \colhead{$L_2$} & \colhead{Ref.} \\ 
			\colhead{} & \colhead{} & \colhead{(days)} & \colhead{} & \colhead{} & \colhead{(K)} & \colhead{(K)} & \colhead{($M_\sun$)} & \colhead{($M_\sun$)} & \colhead{($R_\sun$)} & \colhead{($R_\sun$)} & \colhead{($L_\sun$)} & \colhead{($L_\sun$)} & \colhead{} }

		\startdata
		V692 Pup & A & 0.340  & 1.000  & 0.226  & 5600 & 5398 & 1.06  & 1.06  & 1.00  & 1.00  & 0.88  & 0.76  & (1) \\
		FT UMa & A & 0.655  & 0.984  & 0.153  & 7178 & 7003 & 1.49  & 1.46  & 1.79  & 1.78  & 7.68  & 6.86  & (2) \\
		GU Mon & A & 0.897  & 0.976  & 0.720  & 28000 & 27815 & 8.79  & 8.58  & 4.64  & 4.60  & 11830  & 11324  & (3),(39) \\
		FV CVn & W & 0.315  & 0.969  & 0.069  & 5148 & 5470 & 1.01  & 0.94  & 0.94  & 0.91  & 0.71  & 0.66  & (4) \\
		V796 Cep & A & 0.393  & 0.948  & 0.102  & 6410 & 6403 &  &  &  &  &  &  & (5) \\
		NR Cam & W & 0.256  & 0.942  & 0.006  & 5180 & 5750 & 0.98  & 0.92  & 0.81  & 0.79  & 0.42  & 0.61  & (6) \\
		BF Pav & A & 0.302  & 0.940  & 0.131  & 5201 & 5050 & 0.91  & 0.86  & 0.90  & 0.88  & 0.54  & 0.45  & (7) \\
		BQ Ari & A & 0.282  & 0.910  & 0.341  & 5850 & 5659 & 1.08  & 0.98  & 0.93  & 0.90  & 0.91  & 0.75  & (8) \\
		NSVS 3853195 & A & 0.293  & 0.899  & 0.090  & 5733 & 5637 &  &  &  &  &  &  & (5) \\
		CRTS J213545.6+211104 & W & 0.247  & 0.898  & 0.343  & 4719 & 4839 & 0.87  & 0.78  & 0.83  & 0.79  & 0.31  & 0.31  & (9) \\
		AV Pup & A & 0.435  & 0.896  & 0.101  & 6255 & 6150 & 1.27  & 1.14  & 1.29  & 1.23  & 2.29  & 1.94  & (10) \\
		CSS J015341.9+381641 & A & 0.348  & 0.892  & 0.167  & 5765 & 5657 &  &  &  &  &  &  & (5) \\
		DZ Lyn & A & 0.378  & 0.886  & 0.180  & 6860 & 5068 & 1.25  & 1.11  & 1.18  & 1.12  & 2.82  & 0.76  & (11) \\
		V797 Cep & A & 0.270  & 0.886  & 0.076  & 4833 & 4688 &  &  &  &  &  &  & (5) \\
		YY CMi & A & 1.094  & 0.885  & 0.001  & 6360 & 5710 & 1.25  & 1.12  & 2.32  & 2.20  & 8.13  & 4.67  & (12) \\
		GW Leo & W & 0.336  & 0.881  & 0.028  & 5832 & 6315 & 1.08  & 0.95  & 1.00  & 0.94  & 1.04  & 1.27  & (13) \\
		BG Vul & A & 0.403  & 0.880  & 0.454  & 5868 & 5520 & 1.01  & 0.89  & 1.23  & 1.17  & 1.63  & 1.15  & (14) \\
		1SWASP J210318.76+021002.2 & A & 0.229  & 0.877  & 0.343  & 5850 & 5778 & 1.14  & 1.00  & 1.23  & 1.20  & 1.59  & 1.42  & (15) \\
		IK Boo & A & 0.303  & 0.873  & 0.008  & 5781 & 5422 & 0.99  & 0.86  & 0.91  & 0.85  & 0.64  & 0.73  & (16) \\
		\enddata
		\tablecomments{This table is available in its entirety in machine-readable form in the online version of this article.}
		\tablerefs{(1) \citet{2012BASI...40...51K},
			(2) \citet{2011RAA....11.1158Y},
			(3) \citet{3},
			(4) \citet{2019RAA....19...99M},
			(5) \citet{2017RAA....17...42K},
			(6) \citet{2015NewA...37...64T},
			(7) \citet{2021RAA....21..203P},
			(8) \citet{2021AstL...47..402P},
			(9) \citet{2024PASP..136b4201P},
			(10) \citet{2019arXiv190602466H},
			(11) \citet{2009RAA.....9.1270M},
			(12) \citet{12},
			(13) \citet{2021ARep...65..543P},
			(14) \citet{2014NewA...28...79T},
			(15) \citet{2014NewA...32...10E},
			(16) \citet{2017PASJ...69...62K},
			(17) \citet{2018MNRAS.479.3197A},
			(18) \citet{2020SerAJ.200...19K},
			(19) \citet{2014NewA...28...85E},
			(20) \citet{2005AcA....55..389Z},
			(21) \citet{2018PASA...35....8K},
			(22) \citet{2017AcAau.134..303D},
			(23) \citet{2015MNRAS.448.2890D},
			(24) \citet{2003},
			(25) \citet{2019RAA....19...14K},
			(26) \citet{2011RAA....11.1469K},
			(27) \citet{2014RAA....14.1166S},
			(28) \citet{2016RAA....16..135K},
			(29) \citet{2013AJ....145...39Z},
			(30) \citet{2017NewA...56...10G},
			(31) \citet{2016RAA....16...63J},
			(32) \citet{2022NewA...9701873A},
			(33) \citet{2024NewA..10802162R},
			(34) \citet{2015NewA...34...47E},
			(35) \citet{2011MNRAS.412.1787D},
			(36) \citet{2018NewA...58...90B},
			(37) \citet{2021NewA...8401401K},
			(38) \citet{2013AJ....145....9Y},
			(39) \citet{2019AJ....157..111Y}.
		}
		
	\end{deluxetable*}
\end{longrotatetable}

\bibliography{sample631}
\bibliographystyle{aasjournal}

\end{document}